\documentclass[10pt,twocolumn,twoside] {IEEEtran}
\usepackage[caption=false,font=footnotesize]{subfig}
\usepackage{graphicx}
\usepackage{booktabs}
\usepackage{amsmath}
\usepackage{amssymb}
\usepackage{midfloat}
\usepackage{bm} 
\usepackage{color, cite}
\usepackage{algorithm}
\graphicspath{{figures/}}

\newtheorem{lemma}{Lemma}

\begin{document}
\title{Adaptive Interference Removal for Un-coordinated Radar/Communication Co-existence}
\author{Le Zheng, \emph{Member}, \emph{IEEE}, Marco Lops, \emph{Senior Member} and Xiaodong Wang, \emph{Fellow}, \emph{IEEE}
	\thanks{Le Zheng and Xiaodong Wang are with Electrical Engineering Department, Columbia University, New York, USA, 10027, e-mail: le.zheng.cn@gmail.com, wangx@ee.columbia.edu.
		
	Marco Lops is with the DIEI, Universita degli Studi di Cassino e del Lazio Meridionale, Cassino 03043, Italy (e-mail: lops@unicas.it, e.grossi@unicas.it).}}

\maketitle
\begin{abstract}

Most existing approaches to co-existing communication/radar systems assume that the radar and communication systems are coordinated, i.e.,  they share information, such as relative position, transmitted waveforms and channel state. In this paper, we consider an un-coordinated scenario where a communication receiver is to operate in the presence of a number of radars, of which only a sub-set may be active, which poses the problem of estimating the active waveforms and the relevant parameters thereof, so as to cancel them prior to demodulation. Two algorithms are proposed for such a joint waveform estimation/data demodulation problem, both exploiting sparsity of a proper representation of the interference and of the vector containing the errors of the data block, so as to implement an iterative joint interference removal/data demodulation process. The former algorithm is based on classical on-grid compressed sensing (CS), while the latter forces an atomic norm (AN) constraint: in both cases the radar parameters and the communication demodulation errors can be estimated by solving a convex problem. We also propose a way to improve the efficiency of the AN-based algorithm. The performance of these algorithms are demonstrated through extensive simulations, taking into account a variety of conditions concerning both the interferers and the respective channel states.

\end{abstract}

\section{Introduction}

The ever increasing demand for spectrum and the consequent shortage of available bandwidths pave the way to communication/radar co-existing system architectures \cite{chiriyath2016inner,griffiths2014t09,griffiths2015radar,hessar2016spectrum}. This inevitably produces inter-system interference that degrades the performance of both systems: in particular, the radar transmit power may be large enough to significantly degrade the performance of the communication system. Techniques such as interference mitigation \cite{deng2013interference}, pre-coding or spatial separation \cite{babaei2013practical,aubry2014cognitive,khawar2014spectrum}, waveform design \cite{aubry2014radar,huang2015radar,zheng2017joint,sodagari2012projection} allow both radar and communications to share the spectrum and co-exist: for example, in \cite{babaei2013practical,sodagari2012projection,khawar2014spectrum}, the radar interference is eliminated by forcing the radar waveforms to live in the null space of the interference channel between the radar transmitters and the communication receiver. Motivated by the cooperative methods in cognitive radio networks, later works \cite{lioptimum,li2016mimo,turlapaty2014joint} exploit some prior knowledge to jointly design the radar waveform and communication code-book by minimizing a measure of the mutual interference under certain constraints.

Many existing approaches assume that the radar and communication systems are aware of the existence of each other, and share information: for example, {\em ad hoc} design of radar waveforms or beam-formers is proposed in \cite{deng2013interference,aubry2014radar,babaei2013practical,aubry2014cognitive,huang2015radar} to reduce the mutual interference, while \cite{lioptimum,li2016mimo,turlapaty2014joint,zheng2017joint,sodagari2012projection} rely on making channel information available to the communication and radar through the transmission of pilot training. Otherwise stated, these approaches rely on a centralized architecture, namely on a strict coordination between the active players in order to allow co-existence. In \cite{khawar2014spectrum}, the radar interference is eliminated by forcing the radar waveforms to live in the null space of the interference channel between the radar transmitters and the communication receiver. The algorithm can be extended to the situation with no cooperation between radar and communication by using a blind null space estimation method \cite{manolakos2012blind}. With adaptively adjusted radar waveforms, difficulties for real-time processing arise and some performance loss may happen to the radar.

The situation we refer to in this contribution is one wherein a communication system should share its spectrum with an ensemble of potential interferers, i.e., a set of radar/sensing systems.  We assume full bandwidth overlapping of the active systems, but that not necessarily all of the potentially active radars are transmitting: thus, when observed on a conveniently long time interval, this situation is akin to a highly non-stationary environment, wherein the sources of interference may vary over time, and so do the corresponding waveforms, timing and channels. While the interference produced by the (unique) communication system on the active radars may be neglected, due to both the order of magnitude of the powers in play and some specific countermeasure that can be taken (for example, the active radars may use suitable beam-forming techniques to get rid of interference from a known location \cite{liu2014joint}), the interference produced onto the communication system may be highly detrimental, and must be dealt with. Unlike the anti-jamming in un-coordinated wireless networks \cite{xiao2012jamming,xiao2012mac}, our communication system can be a single TX/RX pair, so it cannot rely on the collaborative broadcasting scheme.

In this un-coordinated scenario, the only information the communication system can rely upon is that the interfering waveforms live in the subspace of a known dictionary, and that they impinge on the communication receiver (RX) with unknown, possibly time-varying delays and coupling coefficients. As a consequence, the communication RX must be made {\em adaptive}, in order to accomplish jointly the two tasks of interference estimation/removal and data demodulation. The approach we propose here focuses on guaranteeing the performance of the communication system and relies on the size of the dictionary from which the radar waveforms are picked up: we first show that, adopting a suitable representation domain,  the interfering signals hitting the communication RX are {\em sparse}. On a parallel track, if iterative demodulation/re-modulation algorithms are implemented, the vector containing the demodulation errors of a data block should be itself sparse (and become sparser and sparser as the iterations go), whereby a joint interference removal/data demodulation process can take great advantage of existing algorithms forcing sparsity constraints.

Unfortunately, however, the scenario we consider in this paper is not ensured to lend itself to direct application of compressed sensing (CS) theory \cite{foucart2013mathematical}, which relies on the fact that signals can be sparsely represented by a finite discrete dictionary \cite{stankovic2013compressive,jokanovic2015reduced,studer2012recovery,yang2014exact}: the presence of relevant continuous parameters, such as delays, could indeed lead to remarkable degradations from model mismatch, should a simple discretization of the parameter space be implemented  \cite{chi2011sensitivity}. We thus also explore the applicability of
 the recently developed mathematical theory of continuous sparse recovery for super-resolution \cite{candes2013super,candes2014towards,tang2013compressed} and in particular of Atomic-Norm (AN) minimization techniques successfully employed for continuous frequency recovery from incomplete data \cite{tang2013compressed,bhaskar2013atomic}, direction-of-arrival estimation \cite{tan2014direction}, interference mitigation \cite{al2016atomic} and line spectral estimation \cite{tang2014robust}.

Given the above framework, we thus propose two algorithms for joint waveform estimation and data demodulation in the overlaid radar/communication architecture, the former based on classical on-grid CS, the latter forcing an AN constraint: in both cases the radar parameters and the communication demodulation errors can be estimated by solving a convex problem. We also propose a way to improve the efficiency of the AN-based algorithm. The merits of these algorithms are demonstrated through extensive simulations, taking into account a variety of conditions concerning both the interferers and the respective channel states.

The remainder of the paper is organized as follows. In Section II, we present the signal model of the co-existed radar and communication system. In Section III, we develop the proposed CS-based algorithms using both the atomic norm and the $\ell_1$-norm. In Section IV, an accelerated algorithm for solving the atomic norm-based algorithm is proposed. Simulation results are presented in Section V. Section VI, finally, contains concluding remarks.

\section{Problem Formulation}

We consider a situation with one communication system and $J$ active radars. Suppose the $j$-th radar transmits the coded waveform
\begin{eqnarray}
s_j(t) = \sum_{n=0}^{N-1} g_j(n) \xi(t - nT),
\end{eqnarray}
where $g_j(n)$ is the $n$-th code, $N$ is the code length, and $\xi(t)$ satisfies the Nyquist criterion with respect to $T$. From now on we assume that such a basic pulse is a Square Root Rased Cosine (SRRC) with excess bandwidth $\beta$, so that the transmit bandwidth is $\frac{1+\beta}{T}$. We also set $N = \tilde N_1 + \tilde N_2$ with $\tilde N_1 = \tilde N_2 = (N-1)/2$ when $N$ is odd and $\tilde N_1 = \tilde N_2 + 1 = N/2$ when $N$ is even. Let $\bm{\bar g}_j = [\bar g_j(- \tilde N_1),\bar g_j(- \tilde N_1 + 1),...,\bar g_j(\tilde N_2)]^T \in \mathbb{C}^{N \times 1}$ be the Discrete Fourier Transform (DFT) of $\bm g_j = [g_j(0),g_j(1),...,g_j(N-1)]^T \in \mathbb{C}^{N \times 1}$, i.e., $\bm{\bar g}_j = \bm F \bm g_j$ with $\bm F = [\bm f_{-\tilde N_1}, \bm f_{-\tilde N_1+1}, ..., \bm f_{\tilde N_2}]^H \in \mathbb{C}^{N \times N}$ denoting the DFT matrix. In practice, the radar waveforms usually have some characteristics to achieve certain performance, so we assume $\bm{\bar g}_j$ lives in a low-dimensional subspace of $\mathbb{C}^{N}$, spanned by the columns of a known $N \times K$ matrix $\bar{\bm D} = [\bm{\bar d}_{-\tilde N_1}, \bm{\bar d}_{-\tilde N_1+1},...,\bm{\bar d}_{\tilde N_2}]^H \in \mathbb{C}^{N \times K}$ with $\bm{\bar d}_n \in \mathbb{C}^{K \times 1}$ and $K \ll N$, i.e., $\bm{\bar g}_j = \bar{\bm D} \bm h_j$ for some unknown $\bm h_j \in \mathbb{C}^{K \times 1}$ such that $\| \bm h_j \|_2 =1$ for $j = 1,2,...,J$. An example of such a situation is waveform diversity, wherein radar systems may vary the transmit waveforms by selecting them in a set - a dictionary - so as to cope with interference, clutter, co-existence with competing systems \cite{hassanien2016dual}. Obviously, if $KJ>N$, the number of unknown variables exceeds $N$, and the problem is for sure infeasible: in general, since the radar waveforms may contain a number of unknown parameters, the condition $K \ll N$ is mathematically necessary, on top of being {\em per se} reasonable.  % Specifically, in the case that the radar waveform is known a priori, we have $K=1$ and $\bar{\bm D}$ becomes an $N \times 1$ vector.

\color{black} For simplicity, we assume that there is one path between the radar TX and the communication RX. This assumption is true for narrow-band radar systems \cite{li2016mimo} or as the interference is dominated by the direct path between the radar TX and the communication RX. The interference produced by $J$ active radars - with $J$ possibly unknown - onto the communication RX can be expressed as
\begin{eqnarray}
\label{eq:yI}
y_I(t) = \sum_{j=1}^{J} \sum_{n=0}^{N-1} c_j g_j(n)\xi(t- n T-\tau_j),
\end{eqnarray}
where $\tau_j$ and $c_j$ denote the delay and complex coupling coefficient of the $j$-th radar, respectively.

We assume the communication TX is not moving and its position is known by the radars, so its effect on the radar can be dealt with via beamforming. The communication TX transmits data symbols $\bm b = [b(0), b(1), ..., b(M-1) ]^T \in \mathbb{C}^{M \times 1}$ with $M \leq N$. Let $\bm b \in {\cal B}$ where $\cal B$ denotes the set of possible $\bm b$ values. Defining the data received at the RX side as $\bm x = [x(0),x(1),...,x(N-1)]^T \in \mathbb{C}^{N \times 1}$, we have
\begin{eqnarray}
\bm x = \bm H \bm A \bm b,
\label{eq:model}
\end{eqnarray}
where $\bm A \in \mathbb{C}^{N \times M}$, and $\bm H \in \mathbb{C}^{N \times N}$ is the channel matrix. This model subsumes a number of communication systems. For example, in a Code-Division Multiple Access (CDMA) system \cite{guo2005randomly}, the elements in $\bm b$ are the symbols transmitted by $M$ active users: The columns of $\bm A$ are the signatures of the users, and $\bm H$ is a diagonal matrix representing the channel gains. Another example is an OFDM system \cite{zhang2015maximum}, in which $\bm A$ is the IDFT matrix with $M = N$, $\bm H$ is the channel matrix and $\bm x$ is the received data in time domain. In practice, $\bm A$ is known and the channel matrix $\bm H$ can be obtained through the transmission of pilot signals. Let $\bm{\bar x} = [\bar x(-\tilde N_1), \bar x(-\tilde N_1 + 1),...,\bar x(\tilde N_2)]^T \in \mathbb{C}^{N \times 1}$ be the DFT of $\bm x$, i.e., $\bm{\bar x} = \bm F \bm x$.

The received communication signal is given by
\begin{eqnarray}
\label{eq:yC}
y_C(t) = \sum_{n=0}^{N-1} x(n) \xi(t - n T - \tau_C),
\end{eqnarray}
where $\tau_C$ denotes the overall delay of the communication transmission. As the communication TX and RX are synchronized, we let $\tau_C=0$ for the convenience of derivation. Notice also that the full bandwidth overlap scenario considered in this paper makes it reasonable to assume that the communication system occupies the full available bandwidth  $\frac{1+\beta}{T}$ and uses the same SRRC basic pulse as the radar systems.

At the communication RX, the signal contains both the communication signal and the radar interference, i.e.,
\begin{eqnarray}
\label{eq:y}
y(t) &=&  y_I(t) + y_C(t) + \tilde w(t), \nonumber \\
&=& \sum_{j=1}^{J} \sum_{n=0}^{N-1} c_j g_j(n)\xi(t-nT-\tau_j) \nonumber \\
&& + \sum_{n=0}^{N-1} x(n) \xi(t - n T) + \tilde w(t),
\end{eqnarray}
where $\tilde w(t)$ is the measurement noise. Projecting $y(t)$ onto $\xi (t - t')$ results in
\begin{eqnarray}
\label{eq:r}
r(t') &=& \left\langle {y(t),\xi (t - t')} \right\rangle \nonumber \\
&=& \sum\limits_{n = 0}^{N-1} {x(n){R_\xi }(t' - n T)}  \nonumber \\
&& + \sum\limits_{j = 1}^J {\sum\limits_{n = 0}^{N-1} {{c_j}{g_j}(n){R_\xi }(t' - n T - {\tau _j})} }  + w(t'),
\end{eqnarray}
where $R_\xi (\cdot)$ is the auto-correlation function of $\xi(\cdot)$, i.e., ${R_\xi }(\tau) = {\left\langle {\xi (t),\xi (t - \tau)} \right\rangle }$ with $\langle \cdot \rangle$ denoting inner product, and $w(t') = \left\langle \tilde w(t) , \xi (t - t') \right\rangle $. The auto-correlation function $R_{\xi}(t)$ is considered substantially time-limited in a finite interval, $[-T',T']$ say: This is a common assumption which is justified by the fact that $R_\xi (t)$ is always vanishingly small for large $t$, and in particular goes to zero as $t^{-3}$ in the considered scenario (see Appendix A). Letting $\tau_{\text{min}}$ and $\tau_{\text{max}}$ be the minimum and the maximum delays, respectively, tied to the corresponding minimal and maximum distances of all of the potential radar systems from the receiver, we define $\bar r(k)$ in \eqref{eq:truebarr}
\begin{figure*}
\begin{eqnarray}
\label{eq:truebarr}
\bar r(k) &=& \int_{\tau_{\text{min}} -T' }^{\tau_{\text{max}} + (N-1)T + T'}  {r(t'){e^{\frac{{ - i2\pi k t'}}{{NT}}}}dt'} \nonumber \\
\label{eq:barr0}
&=& \sum\limits_{n = 0}^{N-1} {x(n) \int_{\tau_{\text{min}} -T' }^{\tau_{\text{max}} + (N-1)T + T'}  {{R_\xi }( t' - nT){e^{\frac{{ - i2\pi k t'}}{{NT}}}} dt'} } + \int_{\tau_{\text{min}} -T' }^{\tau_{\text{max}} + (N-1)T + T'} {w(t'){e^{\frac{{ - i2\pi k t'}}{{NT}}}}dt'} \nonumber \\
&&+ \sum\limits_{j = 1}^J {\sum\limits_{n = 0}^{N-1} {{c_j}{g_j}(n)\int_{ \tau_{\text{min}} - T' }^{ \tau_{\text{max}} + (N-1)T + T'}  {{R_\xi }(t' - n T - {\tau _j} ){e^{\frac{{ - i2\pi k t'}}{{NT}}}}dt'} } } \\
\label{eq:barr}
&\approx& \sum\limits_{n = 0}^{N-1} {x(n)\int_{ - \infty }^{\infty}  {{R_\xi }(t' - nT){e^{\frac{{ - i2\pi k t'}}{{NT}}}}dt'} }  + \int_{\tau_{\text{min}} -T' }^{\tau_{\text{max}} + (N-1)T + T'}  {w(t'){e^{\frac{{ - i2\pi k t'}}{{NT}}}}dt'} \nonumber \\
&& + \sum\limits_{j = 1}^J { {e^{ - i2\pi k  \tau_j' }} \sum\limits_{n = 0}^{N-1} {{c_j}{g_j}(n) \int_{ - \infty }^{ \infty }  {{R_\xi }(t' - n T){e^{\frac{{ - i2\pi k  t'}}{NT}}}dt'} } }
\end{eqnarray}
\end{figure*}
for $k = -\tilde N_1, -\tilde N_1+1,...,\tilde N_2$, where $\tau_j' = \tau_j/NT$. For simplicity, we define $ \bar w(k) = \int_{ \tau_{\text{min}} - T' }^{ \tau_{\text{max}} + (N-1)T + T'}  {w(t'){e^{\frac{{ - i2\pi k t'}}{{NT}}}}dt'}$. It is assumed that $\bar w (k)$ is a complex Gaussian variable with zero mean and variance $\sigma_w^2$, i.e., $\bar w(k) \sim {\cal CN}(0,\sigma_w^2)$. In communication receivers, $\sigma_w^2$ is typically available and can be estimated off-line. Suppose that $R_\xi (\cdot)$ satisfies $\int_{ - \infty }^\infty  {{R_\xi }(t){e^{\frac{{ - i2\pi k t}}{{NT}}}}dt}  \simeq 1$ for $k= -\tilde N_1, -\tilde N_1 +1,..., \tilde N_2$. The above condition is rigorously true for small excess bandwith $\beta$ (in particular, $\beta \leq \frac{1}{N}$), which includes the relevant case that the communication system employs an OFDM format, corresponding to $\beta=0$, while being only approximately true for larger values of $\beta$  \cite{glover2010digital}. Since in an efficient spectrum exploitation context it is mandatory to choose small excess bandwiths, we henceforth assume 
\begin{eqnarray}
\label{eq:integ}
\int_{ - \infty }^\infty  {{R_\xi }(t - nT){e^{\frac{{ - i2\pi k t}}{{NT}}}}dt}  \simeq  e^{\frac{-i2 \pi n k}{N}}.
\end{eqnarray}
Plugging \eqref{eq:integ} into \eqref{eq:barr}, we have 
\begin{eqnarray}
\label{eq:barr2}
\bar r(k) &\approx& \sum\limits_{n = 0}^{N-1} {x(n) e^{\frac{-i2 \pi n k}{N}} }  + \bar w(k) \nonumber \\
&& + \sum\limits_{j = 1}^J { {e^{ - i2\pi k \tau_j'}} \sum\limits_{n = 0}^{N-1} {{c_j}{g_j}(n) e^{\frac{-i2 \pi n k}{N}} } }, \nonumber \\
&=& \bar x(k) + \bar w(k) + \sum\limits_{j = 1}^J {c_j} {\bar g_j}(k) { {e^{- i2\pi k \tau_j' }} },
\end{eqnarray}
where $\bar x(k)$ and $\bar g_j(k)$ are the $k$-th element of $\bm{\bar x}$ and $\bm{\bar g}_j$, respectively.

We define $\bm{\bar r} = [\bar r(-\tilde N_1),\bar r(-\tilde N_1 + 1),...,\bar r(\tilde N_2)]^T \in \mathbb{C}^{N \times 1}$. As outlined in the introduction, the communication RX has to remove the radar interference from the measurement with no knowledge of the delays and the waveforms of the active radars, i.e., under uncertainty concerning $\{ \tau_j' \}_{1 \leq j \leq J}$ and $\{ \bm{\bar g}_j \}_{1 \leq j \leq J}$. In principle, demodulation may be undertaken simply ignoring the presence of interference, i.e., through the operation $\bm{\hat b} = \Psi(\bm{\bar r})$, with $\Psi(\cdot)$ the decoding function operating on the received signal, which would obviously lead to an uncontrolled symbol-error-rate (SER). The approach we take here instead relies on a joint interference-estimation symbol-demodulation process. In particular, defining $\bm{\hat x} = [\hat x(0), \hat x(1),...,\hat x(N-1)]^T = \bm{H} \bm A \bm{\hat b} \in \mathbb{C}^{N \times 1}$ as the ``estimated communication signal", the presence of errors in the decision process results into a non-zero difference vector $\bm{\bar x} - \bm F \bm{\hat x} = \bm F \bm H \bm A \bm v$ where $\bm v = \bm b - \bm{\hat b}$. Obviously, the $k$-th element of $\bm z = [z(-\tilde N_1),z(-\tilde N_1 + 1),...,z(\tilde N_2)]^T = \bm{\bar r} - \bm{\hat x}$ is given by
\begin{eqnarray}
\label{eq:z}
z(k) &=& \left\langle \bm H \bm A \bm v, \bm f_k \right\rangle + \sum\limits_{j = 1}^J {{c_j}{{\bar g}_j}(k){e^{-i2\pi k{\tau_j'}}}} + \bar w(k), \nonumber \\
&=& \left\langle \bm H \bm A \bm v, \bm f_k \right\rangle + \sum\limits_{j = 1}^J {c_j} \bm a(\tau_j')^H \bm e_k \bm{\bar d}_k^H \bm h_j + \bar w(k), \nonumber \\
&=& \left\langle \bm H \bm A \bm v, \bm f_k \right\rangle + \left\langle \bm X , \bm{\bar d}_k \bm e_k^H \right\rangle + \bar w(k),
\end{eqnarray}
where we have defined $\left\langle \bm X, \bm Y \right\rangle = \text{Tr}(\bm Y^H \bm X)$, $\bm X = \sum_{j=1}^{J} c_j \bm h_j \bm a(\tau_j')^H$ with $\bm a(\tau) = [e^{- i2 \pi \tilde N_1 \tau}, e^{i2 \pi (-\tilde N_1 + 1) \tau},  ...,e^{i2\pi \tilde N_2 \tau}]^T$, and $\bm e_k$ the $(\tilde N_1 + k +1)$-th column of the $N \times N$ identity matrix $\bm I_N$. Once estimates of $\bm X$ and $\bm v$ are available, the radar interference can be obtained and canceled from the measurements and the symbols re-demodulated. Hence, the main problem is to estimate $\bm X$ and $\bm v$ from the noisy measurements $\bm z$. Notice that $z(k)$ contains both the radar interference and the residual of communication signal caused by the mis-demodulations. The mixing of both signals causes great difficulties for the estimation, which inspires us to exploit some structural information about the desired solution, and in particular sparsity, as detailed in the next section.

\section{Proposed Algorithms}

Equation \eqref{eq:z} highlights that data demodulation and interference mitigation are {\em coupled}, in the sense that they should be accomplished jointly and that poor performance in estimating either one has detrimental effects on the estimate of the other. In fact, in order to remove the radar interference, we need to estimate the matrix $\bm X$ from the observations \eqref{eq:z}, but this would obviously require also estimating the error vector $\bm v$, which boils down to correctly demodulating the data block. To this end, we design iterative algorithms, exploiting structural information on the desired solution.  In the $l$-th iteration, the demodulated symbols are denoted as $\bm{\hat b}^{(l)}$, and $\bm z^{(l)} = \bm{\bar r} - {\bm F} \bm{\hat x}^{(l-1)}$ where $\bm{\hat x}^{(l-1)} = \bm H \bm A \bm{\hat b}^{(l-1)}$ is the estimate of $\bm x$. \color{black} The following two types of sparsity are exploited in the problem:
\begin{enumerate}
	\item  The signal $\bm X$ is a combination of $J$ complex exponentials $\bm a(\tau_j')^H$ with unknown modulation $c_j \bm h_j$, and the number of complex exponentials is much smaller than the dimension of $\bm z$, i.e., $J \ll N$.
	\item Ideally, the vector $\bm v$ should be an all-zero vector. As a consequence, denoting $\bm v^{(l)} = \bm b - \bm{\hat b}^{(l-1)}$ the result of the $l$-th iteration, we want $\|\bm v^{(l)}\|_0 \triangleq L_l$ to be as small as possible, and in any case we want to force the condition $L_l \ll N$.
\end{enumerate}

\subsection{Joint Waveform Estimation and Demodulation Based on On-grid CS Algorithm}

In the first iteration, $\bm{\hat b}^{(0)} = \Psi(\bm{\bar r})$ so $\bm v^{(1)} = \bm b - \bm{\hat b}^{(0)}$. According to \eqref{eq:z}, jointly estimating $c_j \bm h_j$, $\bm a(\tau_j')$ and $\bm v^{(1)}$ is a non-linear problem. However, it can be linerized by using an overcomplete dictionary matrix
\begin{eqnarray}
	\tilde{\bm A}= [\bm a(\tilde \tau_1'), \bm a(\tilde \tau_2'),..., \bm a(\tilde \tau_{\tilde J}')] \in \mathbb{C}^{N \times \tilde J},
\end{eqnarray}
with $\{\tilde \tau_j' \}_{j = 1, 2,..., \tilde J}$ denoting sets of uniformly spaced points of the radar delays. Define $\tilde J$ as the number of columns of $\tilde{\bm A}$ where $\tilde J \geq N$. For sufficiently large $\tilde J$, the delay is densely sampled. Let $\bm{\alpha} = [ \tilde c_{1} \bm{\tilde h}_{1}^T, \tilde c_{2} \bm{\tilde h}_{2}^T, ... , \tilde c_{\tilde J} \bm{\tilde h}_{\tilde J}^T ]^T \in \mathbb{C}^{\tilde J K \times 1}$ be the sparse vector whose non-zero elements correspond to $c_j \bm h_j$ in \eqref{eq:z}.

As usual, forcing a constraint onto the $\ell_0$-norm is impractical, since it results in an  NP-hard non-convex optimization problem, and $\ell_1$-norm regularization is used instead, i.e., $\| \bm{\alpha} \|_1 = \sum_{k= 0}^{\tilde J K-1} |\alpha(k)|$ and $\|\bm{v}\|_1 = \sum_{k=0}^{M-1} |v(k)|$. Define ${\cal V}$ as the set of all possible differences $\bm{b}- \bm{\hat b}^{(0)}$ when the two vectors both belong to ${\cal B}$. We notice that the constraint $\bm v^{(1)} \in {\cal V}$ results in a non-convex problem, and would cause much difficulty in solving the optimization problem. This constraint is relaxed and the non-linear joint estimation problem is thus reduced to a linear parameter estimation problem, i.e., the estimation of the linear amplitude vectors $\bm{\alpha}$ and $\bm{v}^{(1)}$, under a sparsity constraint:
\begin{eqnarray}
\label{eq:cs}
(\bm{\hat \alpha}^{(1)}, \bm{\hat v}^{(1)}) &=& \arg\min_{\bm{\alpha} \in \mathbb{C}^{\tilde J K \times 1}, \bm{v}^{(1)} \in \mathbb{C}^{N \times 1}}  \frac{1}{2} {\left\| \bm z - \bm \Phi \bm v^{(1)} - \bm \Upsilon \bm{\alpha} \right\|_2^{2}} \nonumber \\
&& + \tilde \lambda \| \bm{\alpha} \|_1 + \tilde \gamma \| \bm{v}^{(1)} \|_1,
\end{eqnarray}
where
\begin{eqnarray}
\bm \Phi = \left[ {\begin{array}{*{20}{c}}
	{ \bm f_{-\tilde N_1}^H \bm H \bm A }\\
	{ \bm f_{-\tilde N_1+1}^H \bm H \bm A }\\
	\vdots \\
	{ \bm f_{\tilde N_2}^H \bm H \bm A }
	\end{array}} \right],
\end{eqnarray}
and $\bm \Upsilon$ is given by \eqref{eq:Upsilon}.
\begin{figure*}
\begin{eqnarray}
\label{eq:Upsilon}
\bm \Upsilon  = \left[ {\begin{array}{*{20}{c}}
	{\bm a{{({{\tilde \tau}_1'})}^H}{\bm e_{-\tilde N_1}} \bm{\bar d}_{-\tilde N_1}^H}&{\bm a{{({{\tilde \tau }_2'})}^H}{\bm e_{-\tilde N_1}} \bm{\bar d}_{-\tilde N_1}^H}& \cdots &{\bm a{{({{\tilde \tau }_{\tilde J}'})}^H}{\bm e_{-\tilde N_1}} \bm{\bar d}_{-\tilde N_1}^H}\\
	{\bm a{{({{\tilde \tau }_1'})}^H}{\bm e_{-\tilde N_1 + 1}} \bm{\bar d}_{-\tilde N_1 + 1}^H}&{\bm a{{({{\tilde \tau }_{-\tilde N_1 + 1}'})}^H}{\bm e_{-\tilde N_1 + 1}} \bm{\bar d}_{-\tilde N_1 +1}^H}& \cdots &{\bm a{{({{\tilde \tau }_{\tilde J}'})}^H}{\bm e_{-\tilde N_1 +1}} \bm{\bar d}_{-\tilde N_1 +1}^H}\\
	\vdots & \vdots & \ddots & \vdots \\
	{\bm a{{({{\tilde \tau }_1'})}^H}{\bm e_{\tilde N_2}} \bm{\bar d}_{\tilde N_2}^H}&{\bm a{{({{\tilde \tau }_2'})}^H}{\bm e_{\tilde N_2}} \bm{\bar d}_{\tilde N_2}^H}& \cdots &{\bm a{{({{\tilde \tau }_{\tilde J}'})}^H}{\bm e_{\tilde N_2}} \bm{\bar d}_{\tilde N_2}^H}
	\end{array}} \right].
\end{eqnarray}
\end{figure*}
The parameters $\tilde \lambda$ and $\tilde \gamma$ are weights determining the sparsity of the reconstruction. In practice, we set $\tilde \lambda, \tilde \gamma \backsimeq \sigma_w \sqrt{2 \log(\tilde J K)}$. As \eqref{eq:cs} is convex, it can be solved with standard convex solvers. The computational complexity of problem solving depends on the dimension of $\bm \alpha$ and $\bm v$. Specifically, if \eqref{eq:cs} is solved by using the interior point method \cite{boyd2004convex}, the complexity in each iteration is ${\cal O}((\tilde J K + N)^3)$.

By solving \eqref{eq:cs}, we obtain the estimates $\bm{\hat \alpha}^{(1)}$ and $\bm{\hat v}^{(1)}$, whereby the symbols can be corrected and re-demodulated as:
\begin{eqnarray}
\label{eq:estb}
\bm{\hat b}^{(1)} = \arg\min_{\bm b \in {\cal B}} \left\| \bm b - \bm{\hat b}^{(0)} - \bm{\hat v}^{(1)} \right\|_2.
\end{eqnarray}
Notice that the re-demodulation process makes use of the structural information of the communication symbols, i.e., $\bm b \in {\cal B}$. After the re-demodulation, the demodulated symbols belong to the constellation alphabet, whereby  $\bm{v}^{(2)} \in {\cal V}$.

When the estimates of the symbols in the first iteration is accurate and the measurement noise is small, i.e., $\| \bm{v}^{(1)} \|_0$ and $\sigma_w$ are both small, then all the mistakenly demodulated symbols can be corrected by applying \eqref{eq:estb}. In many cases, however, the interference from the radars is strong, and $\bm{\hat b}^{(0)}$ contains many demodulation errors. Thus, we need to iterate the joint interference removal/data demodulation process. Specifically, in the $l$-th iteration ($l \geq 2$), $\bm{\hat \alpha}^{(l)}$ and $\bm{\hat v}^{(l)}$ are estimated as
\begin{eqnarray}
\label{eq:cs2}
(\bm{\hat \alpha}^{(l)}, \bm{\hat v}^{(l)}) &=& \arg\min_{\bm{\alpha} \in \mathbb{C}^{\tilde J K \times 1}, \bm{v}^{(l)} \in \mathbb{C}^{N \times 1}}  \tilde \lambda \| \bm{\alpha} \|_1 + \tilde \gamma \| \bm{v}^{(l)} \|_1 \nonumber \\
&& + \frac{1}{2} {\left\| \bm z^{(l)} - \bm \Phi \bm v^{(l)} - \bm \Upsilon \bm{\alpha} \right\|_2^{2}}.
\end{eqnarray}
Then re-demodulation is undertaken by \eqref{eq:estb} with $\bm{\hat b}^{(0)}$ replaced by $\bm{\hat b}^{(l-1)}$ and $\bm{\hat v}^{(1)}$ replaced by $\bm{\hat v}^{(l)}$. As the iteration goes, some wrong symbols are corrected, and the demodulation error $\bm v^{(l)}$ becomes sparser as $l$ increases. The proposed algorithm iterates until $\bm{\hat b}^{(l-1)} = \bm{\hat b}^{(l)}$ or the maximum number of iterations is reached.

Let $\bm{\hat \alpha}$ be the estimate of $\bm \alpha$ when the algorithm terminates. The radar delays can be identified by locating the non-zero entries of $\bm{\hat \alpha}$. If the solution $[\hat \alpha_{(j-1)K}, \hat \alpha_{(j-1)K+1},..., \hat \alpha_{(j-1)K +K -1}]^T$ is either non-zero or has elements larger than a pre-set threshold, i.e., $\| \tilde c_j \bm{\tilde h}_j \|_2 \neq 0$, then a radar interference exists at delay $\tilde \tau_j'T$. Notice that one cannot resolve the inherent scaling ambiguity between each $\tilde c_j$ and the corresponding $\bm{\tilde h}_j$, which is in any case not essential since it is the product $c_j \bm{g}_j$ the measure of interest for radar interference removal purposes. The estimated time domain radar waveform is then given by
\begin{eqnarray}
\label{eq:hatg}
\hat c_j \bm{\hat g}_j = \bm F^{H} \bar{\bm D} \tilde c_j \bm{\tilde h}_j.
\end{eqnarray}

The CS algorithm based on $\ell_1$-minimization (CS-L1) is capable of super-resolving the spectrum of the sparse signal under certain conditions of the matrices $\bm \Phi$ and $\bm \Upsilon$ \cite{eldar2009robust}. For comparison purposes, and to quantify, at the performance assessment stage, the loss induced by the uncertainty on the active radars positions, it is worth exploring also the case that the delays of such active radar systems are known at the communication receiver. In such a simplified scenario, which assumes that training pilot signals are periodically transmitted for channel sensing purposes, a modified version of the proposed algorithm can be easily derived, by using arguments similar to those employed above. Indeed, since the radar delays $\tau_j'$ are known precisely, we let $\bar{\bm \alpha} = [c_1 \bm h_1^T, c_2 \bm h_2^T, ..., c_J \bm h_J^T]^T \in \mathbb{C}^{KJ \times 1}$, and \eqref{eq:z} can be re-written as
\begin{eqnarray}
\label{eq:z2}
z(k) = \left\langle \bm H \bm A \bm v, \bm f_k \right\rangle + \bm \phi_k \bar{\bm \alpha} + \bar w(k),
\end{eqnarray}
where $\bm \phi_k = [\bm a(\tau_1')^H \bm e_k \bm{\bar d}_k^H, \bm a(\tau_2')^H \bm e_k \bm{\bar d}_k^H, ..., \bm a(\tau_J')^H \bm e_k \bm{\bar d}_k^H] \in \mathbb{C}^{1 \times KJ}$. The algorithm with known timing and radar delay has the same procedure as that of CS-L1. Notice that $\bar{\bm \alpha}$ is not sparse. In the $l$-th iteration, $\bar{\bm \alpha}$ and the demodulation error $\bm{v}^{(l)}$ can be estimated by solving \eqref{eq:cs2} with $\bm \Phi = [\bm \phi_{-\tilde N_1}^T,\bm \phi_{-\tilde N_1+1}^T,...,\bm \phi_{\tilde N_2}^T]^T \in \mathbb{C}^{N \times KJ}$, $\tilde \lambda \| \bm{\alpha} \|_1$ removed and $\bm{\alpha} \in \mathbb{C}^{\tilde J K \times 1}$ replaced by $\bar{\bm \alpha} \in \mathbb{C}^{JK \times 1}$.

\subsection{Joint Waveform Estimation and Demodulation Based on Off-grid CS Algorithm}

As anticipated, \eqref{eq:z} reveals that $\bm X$ is a linear combination of modulated complex exponentials with arbitrary phases, which do not in general correspond to the point of a discrete grid: the off-grid radar position can lead to mismatches in the model and deteriorate the performance. In this subsection, we use the atomic norm to build a sparse representation which does not suffer from the off-grid problem. We define the atomic norm \cite{yang2016super} associated to $\bm X$ as
\begin{eqnarray}
\| \bm X \|_{\cal A} &=& \inf \left\{ \mu>0: \bm X \in \mu \text{conv}({\cal A}) \right\} \nonumber \\
&=& \inf_{c_j, \tau_j', \|\bm h_j\|_2 = 1} \left\{ \sum_j |c_j|: \bm X = \sum_j c_j \bm h_j \bm a(\tau_j')^H \right\}, \nonumber \\
\end{eqnarray}
where $\text{conv}(\cdot)$ denotes the convex hull of the input atom set, and the set of atoms is defined as
\begin{eqnarray}
{\cal A} = \left\{ \bm h \bm a(\tau')^H: \tau' \in [0,1), \|\bm h\|_2 = 1, \bm h \in \mathbb{C}^{K \times 1} \right\}.
\end{eqnarray}
For future developments, we introduce the following equivalent form of the atomic norm for the atom set $\cal A$ \cite{yang2016super}:
\begin{eqnarray}
\label{eq:atomic}
\| \bm{X} \|_{\cal A} = \mathop {\inf}\limits_{\bm u, \bm T} \left\{ \begin{array}{l}
\frac{1}{2N}{\rm{Tr}}({\rm{Toep}}(\bm u)) + \frac{1}{2} {\rm Tr}(\bm T),\\
{\rm s.t.} \left[ {\begin{array}{*{20}{c}}
	{{\rm{Toep}}(\bm u)}& \bm{X}^H\\
	{{\bm{X}}}& {\bm T}
	\end{array}} \right] \succeq 0
\end{array} \right\},
\end{eqnarray}
where $\bm u \in \mathbb{C}^{N \times 1}$ is a complex vector whose first entry is real, ${\rm Toep}(\bm u)$ denotes the $N \times N$ Hermitian Toeplitz matrix whose first column is $\bm u$, and $\bm T$ is a Hermitian $K \times K$ matrix.

Based on \eqref{eq:z}, and paralleling the arguments outlined in the previous sub-section, the $l$-th iteration achieves estimates, $\bm{\hat X}^{(l)}$ and $\bm{\hat v}^{(l)}$ say, of $\bm X$ and $\bm v^{(l)}$ by processing $\bm z^{(l)}$ and solving the optimization problem:
\begin{eqnarray}
\label{eq:atomicnorm}
&& (\bm{\hat X}^{(l)}, \bm{\hat v}^{(l)}) = \mathop {\min }\limits_{\bm{X},\bm{v}^{(l)}} \lambda \| \bm{X} \|_{\cal A} + \gamma \| \bm{v}^{(l)} \|_1 \nonumber \\ 
&& + \sum_{k=- \tilde N_1}^{\tilde N_2} \frac{1}{2} {\left| z^{(l)}(k) - \left\langle \bm f_k , \bm H \bm A \bm v^{(l)} \right\rangle - \left\langle \bm X , \bm{\bar d}_k \bm e_k^H \right\rangle \right|^{2}}, \nonumber \\
\end{eqnarray}
where $\lambda > 0$ and $\gamma>0$ are the weight factors. In practice, we set $\lambda \backsimeq \sigma_w \sqrt{KN \log(KN)}$ and $\gamma \backsimeq \sigma_w \sqrt{K \log(KN)}$. In the light of \eqref{eq:atomic}, the above can be transformed into the following Semi-Definite Programming (SDP):
\begin{eqnarray}
\label{eq:SDP}
&& (\bm{\hat X}^{(l)}, \bm{\hat v}^{(l)}) = \arg \min_{\bm{X}, \bm T, \bm u, \bm{v}^{(l)}} \frac{\lambda}{2N} {\text{Tr}}\left( {\rm Toep}(\bm u) \right) \nonumber \\
&& + \sum_{k=-\tilde N_1}^{\tilde N_2} \frac{1}{2} {\left| z^{(l)}(k) - \left\langle \bm f_k , \bm H \bm A \bm{v}^{(l)} \right\rangle - \left\langle \bm X , \bm{\bar d}_k \bm e_k^H \right\rangle \right|^{2}} \nonumber  \\
&& + \frac{\lambda \text{Tr}(\bm T)}{2} + \gamma \|\bm{v}^{(l)}\|_1, \\
 && \text{s.t.}\left[ {\begin{array}{*{20}{c}}
	{{\rm Toep}(\bm u)}& \bm{X}^H \\
	{{\bm{X}}}& \bm T
	\end{array}} \right] \succeq 0, \nonumber
\end{eqnarray}
where $\text{Toep}(\cdot)$ denotes the Toeplitz matrix whose first column is the input vector. The above problem is convex, and can be solved by using a convex solver. The corresponding computational load is ${\cal O}((N + K)^6)$ per iteration if the interior point method is applied. We name the algorithm based on solving \eqref{eq:atomicnorm} as the CS Atomic-Norm (CS-AN) based algorithm. Similar to the CS-L1 algorithm, CS-AN iterates \eqref{eq:SDP} and \eqref{eq:estb} until $\bm{\hat b}^{(l-1)} = \bm{\hat b}^{(l)}$. From now on,  $\bm{\hat X}$ denotes the estimate of $\bm X$ when the algorithm terminates.

We hasten to underline here that the atomic norm can be exploited to enforce sparsity in the continuous domain $\cal A$ without any discretization \cite{tang2013compressed,yang2016super}, thus leading to better estimation performance compared to on-grid CS techniques. Atomic-norm-based optimization has also been shown to achieve better de-noising performance than subspace methods such as MUSIC \cite{bhaskar2013atomic}. Notice that if $\bm X$ has low rank, it is amenable to nuclear-norm-based estimation. However, the nuclear norm-based approach cannot fully take advantage of the signal structure (e.g., in our scenario $\bm X$ is a combination of multiple complex exponentials $\bm a(\tau_j')^H$ with unknown modulation $c_j \bm h_j$), whereby its performance has been shown to be inferior to that of atomic norm-based optimization \cite{fernandez2016demixing,chen2014robust}.

Solving \eqref{eq:SDP} does not directly provide estimates of the delays of the active radars, $\{ \tau_j'\}_{j=1}^J$. Notice however that each row of $\bm{\hat X}$ is a linear combination of several complex exponentials, in that, denoting $\bm{\hat X}_{k,1:N} \in \mathbb{C}^{1 \times N}$ the $k$-th row of $\bm{\hat X}$, we have $\bm{\hat X}_{k,1:N} = \sum_{j=1}^{\hat J} \hat c_j \hat h_j(k) \bm a(\hat \tau_j')^H$. Hence, $\hat \beta_j(k) = \hat c_j \hat h_j(k)$ and $\hat \tau_j'$ can be obtained by  MUSIC \cite{naha2015determining} or prony's method \cite{stoica1997introduction} with $\bm{\hat X}_{k,1:N}$ as input. Denoting $\text{MUSIC}(\cdot)$ the operation of the MUSIC algorithm, it outputs $\hat J_k$ components with different amplitudes and delays\footnote{Here $\hat J_k$ is not necessarily the same as $\hat J$ because $\beta_j(k)$ can be zero while $\beta_j(l) \neq 0$ for some $l \neq k$. For example, the radar has $K$ candidate waveforms, and select one for transmission. In such case, $\bm h_j$ is a $K \times 1$ vector having $K-1$ zero elements and one element with magnitude 1. As a result, the number of output delays should satisfy $\tilde J_k \leq \hat J$.}, i.e.,
\begin{eqnarray}
{\cal T}_k = \{ \bar \beta_j(k), \bar \tau_j'(k) \}_{j=1,2,..., \bar J_k} = \text{MUSIC}(\bm{\hat X}_{k,1:N})
\end{eqnarray}
for $k = 1,2,...,K$. An association procedure is thus needed to combine the different $\bar \tau_j'(k)$s and come up with the estimates, $\bar \tau_j'$ say, of the radar signals delays. In practice, the calculation of $\bar \tau_j'$ may not be accurate due to computational errors. Hence, if the estimated delays for different $k$ are closer than a small threshold $\delta$, then they are regarded as generated by the same radar, and the corresponding $\bar \tau_j'(k)$s are combined. For clarity, we summarize the association process in Algorithm 1, where ${\cal S} = \{(\hat \tau_{j}', \hat c_{j} \bm{\hat h}_{j}) \}_{j=1,2,...,\hat J}$ is the set of the estimated radar delays and waveform parameters. In the algorithm, for $\{ \bar \beta_l(k), \bar \tau_l'(k) \}$, if there exists $(\hat \tau_m', \hat c_m \bm{\hat h}_m) \in {\cal S}$ such that $|\hat \tau_m' - \bar \tau_l'(k)| \leq \delta$ and $\hat c_m \hat h_m(k) = 0$, then both components belongs to the same radar and the radar delay is updated by the weighted summation of $\bar \tau_l'(k)$ and $\hat \tau_m'$. Otherwise, an additional radar with the estimated delay and amplitude is added to the set $\cal S$. Notice that, once again, the inherent scaling ambiguity between each $\hat c_j$ and the corresponding $\bm{\hat h}_j$. Hence, we only estimate $c_j \bm{g}_j$ via \eqref{eq:hatg} with $\tilde c_j \bm{\tilde h}_j$ replaced by $\hat c_j \bm{\hat h}_j$.
\begin{algorithm}[htbp]
	\label{tab:A0}
	\caption{Radar delay and path gain estimation}
	\begin{tabular}{lcl}
		Input ${\cal T}_k$ for $k=1,2,...,K$, $\delta$.\\
		1, $\hat J = 0$, ${\cal S} = \{ \}$. \\
		\sf{For $k=1,2 ,..., K$} \\
		\hspace{0.4cm} \sf{For $l=1,2 ,..., \hat J_k$}\\
		\hspace{0.8cm} \sf{If there exists $(\hat \tau_m', \hat c_m \bm{\hat h}_m) \in {\cal S}$ such that} \\
		\hspace{0.8cm} \sf{$|\hat \tau_m' - \bar \tau_l'(k)| \leq \delta$ and $\hat c_m \hat h_m(k) = 0$} \\
		\hspace{1.2cm} 2, $\hat \tau_m' = \frac{\sum_{j=1}^K |\hat c_m {\hat h}_m(j)| \hat \tau_j' + | \bar \beta_l(k) | \bar \tau_l'(k)}{\sum_{j=1}^K |\hat c_m {\hat h}_m(j)| + | \bar \beta_l(k) |}$. \\
		\hspace{1.2cm} 3, $\hat c_m \hat h_m(k) = \bar \beta_l(k)$. \\
		\hspace{0.8cm} \sf{Else} \\
		\hspace{1.2cm} 4, ${\cal S} = \{ {\cal S}, (\hat \tau_{\hat J}', \hat c_{\hat J} \bm{\hat h}_{\hat J}) \}$ where $\hat c_{\hat J} \hat h_{\hat J}(k) = \bar \beta_l(k)$ \\
		\hspace{1.5cm} and $\hat c_{\hat J} \hat h_{\hat J}(m) = 0$ for $m \neq k$. \\
		\hspace{1.2cm} 5, $\hat J = \hat J+1$.\\
		\hspace{0.8cm} \sf{End If} \\
		\hspace{0.4cm} \sf{End For} \\
		\sf{End For} \\
		\midrule
		Return $\cal S$, $\hat J$.\\
	\end{tabular}
\end{algorithm}

An alternative approach to delay estimation consists in solving the dual problem of \eqref{eq:atomicnorm}, which is given by
\begin{eqnarray}
\label{eq:dual}
\max_{\bm \nu} &&{\left\langle {\bm \nu,\bm{z}^{(l)}} \right\rangle _{\mathbb R}} - \frac{1}{2}\left\| {\bm \nu} \right\|_2^2, \\
\text{s.t.} && \| {\cal D}(\bm \nu) \|_{\cal A}^* \leq \lambda \nonumber, \\
&& \left\| \sum_{k=1}^{N} \nu_k \bm A^H \bm H^H \bm f_k \right\|_\infty \leq \gamma \nonumber,
\end{eqnarray}
where $\bm \nu \in \mathbb{C}^{N \times 1}$ is the dual variable, and $\| {\cal D}(\bm \nu) \|_{\cal A}^* = \sup_{\| \bm X \|_{\cal A} \leq 1} \langle {\cal D}(\bm \nu) , \bm X \rangle_{\mathbb R}$ is the dual norm with ${\cal D}(\bm \nu) = \sum_n \nu(n) \bm{\bar d}_n \bm e_n^H \in \mathbb{C}^{K \times N}$, $\langle {\cal D}(\bm \nu) , \bm X \rangle_{\mathbb R} = \text{Re}(\text{Tr}(\bm X^H {\cal D}(\bm \nu)))$. Following the derivation in \cite{yang2016super,fernandez2016demixing}, the following lemma can be obtained:
\begin{lemma}
	Suppose $\bm{\hat X} = \sum_{j=1}^{\hat J} \hat c_j \bm{\hat h}_j \bm a(\hat \tau_j')^H$ and $\bm{\hat v}$ are the primal solutions, then the dual polynomial $\bm q(\tau')= {\cal D}(\bm \nu) \bm a(\tau')$ satisfies
	\begin{eqnarray}
	\label{eq:dualc}
	&& \bm q(\hat \tau') = \lambda \frac{\hat c_j}{|\hat c_j|} \bm{\hat h}_j, j=1,2,...,\hat J,\\
	&& \left( \sum_{k= -\tilde N_1}^{\tilde N_2} \nu_k \bm A^H \bm H^H \bm f_k \right)_j = \gamma \frac{\hat v_j^{(l)}}{|\hat v_j^{(l)}|}, \nonumber \\
	&& \forall \hat v_j^{(l)} \ne 0, j=0,1,...,M-1,
	\end{eqnarray}
	where $\hat J$ is the number of estimated delays, and $\left( \cdot \right)_j$ denotes the $j$-th element of the input vector.
\end{lemma}

Based on \eqref{eq:dualc}, the delays of the active radars can be obtained by identifying points where the dual polynomial has modulus $\lambda$, i.e., $\|\bm q(\hat \tau')\|_2 = \lambda$. Moreover, the dual solution provides another way to detect the mistaken demodulation: in places where mistaken demodulation occurs, the magnitude of $\sum_{k=-\tilde N_1}^{\tilde N_2} \nu_k \bm A^H \bm H^H \bm f_k$ equals $\gamma$.

It is worth underlining that not only does the proposed algorithm correct the demodulation error, but it also provides estimates of $\{\tau_j'\}_{j=1}^J$ and 
$\{ c_j \bm{h}_j\}_{j=1}^J$ . Hence, without requiring the radar to transmit pilots, the communication system is able to estimate the channel state generated by  active radar TX's. 

% In some situations . On this basis and optimize its waveform to reduce the interference on the radar. Based on the estimated radar waveform and location, the communication waveform can be optimized to reduce its interference on the radar.

%\subsection{Demodulation}

% As is shown in the previous section, the measurements of the communication RX contains the radar interference, from which the demodulation performance may suffer. In this subsection, an iterative algorithm is proposed for the joint demodulation and waveform estimation.

% For clarity, we summarize the proposed CS-AN algorithm in Algorithm 2. For the CS-L1 algorithm, line 4 becomes estimating $\bm{\hat \alpha}^{(l)}$ and $\bm{\hat v}^{(l)}$ by solving \eqref{eq:cs} with $\bm z$ replaced by $\bm z^{(l)}$. Line 6 becomes estimating $\bm{\hat \alpha}$ by solving \eqref{eq:cs} with $\bm z$ replaced by $\bm{\bar r} - \bm F \bm H \bm A \bm b^{(l)}$. For the estimation of radar waveform, line 7 and line 8 are removed and line 9 becomes estimating $\hat c_j \bm{\hat g}_j$ via \eqref{eq:hatg}.

\subsection{Example}
The previous discussion highlights that the inherent coupling of interference estimation and data demodulation has a deep impact on the performance of the proposed approach. In order to illustrate further this point and to highlight the rationale behind the AN criterion, we consider a simple scenario wherein an OFDM system with $N=65$ and ${\bm F} {\bm H} {\bm A} = {\bm I}_N$ is to co-exist with $J=2$ active systems out of a set of $K=3$ radars. In the simulations, $\bar r(k)$ is generated according to \eqref{eq:truebarr} with $T' = 2N T$.

We focus on the results of the first iteration of \eqref{eq:cs} and \eqref{eq:SDP} in order to assess the ability of the CS-L1 and the CS-AN algorithms of detecting, identifying and ranging the active transmitters. The simulations have been performed by generating data according to \eqref{eq:z}, while the $\bm h_j$'s are uniformly generated with $\|\bm h_j\|_2 = 1$ for $j=1,2,...,J$. Due to the coupling of interference estimation and data demodulation, the initial symbol error rate (SER) plays a key role, and we assume the two initial values of 0.1 and 0.3. In the example, the communication system uses binary phase-shift keying (BPSK) and the wrong symbols are randomly placed in $\bm{\hat b}^{(0)}$ based on the SER. The grid parameter of the CS algorithm has been set as $\tilde J = 4 N$. For both algorithms, the weights have been optimized so that the mean-squared-error (MSE) of the estimate is minimized.

\begin{figure}
	\centering
		
	\subfloat[][]{\includegraphics[width=1.8in]{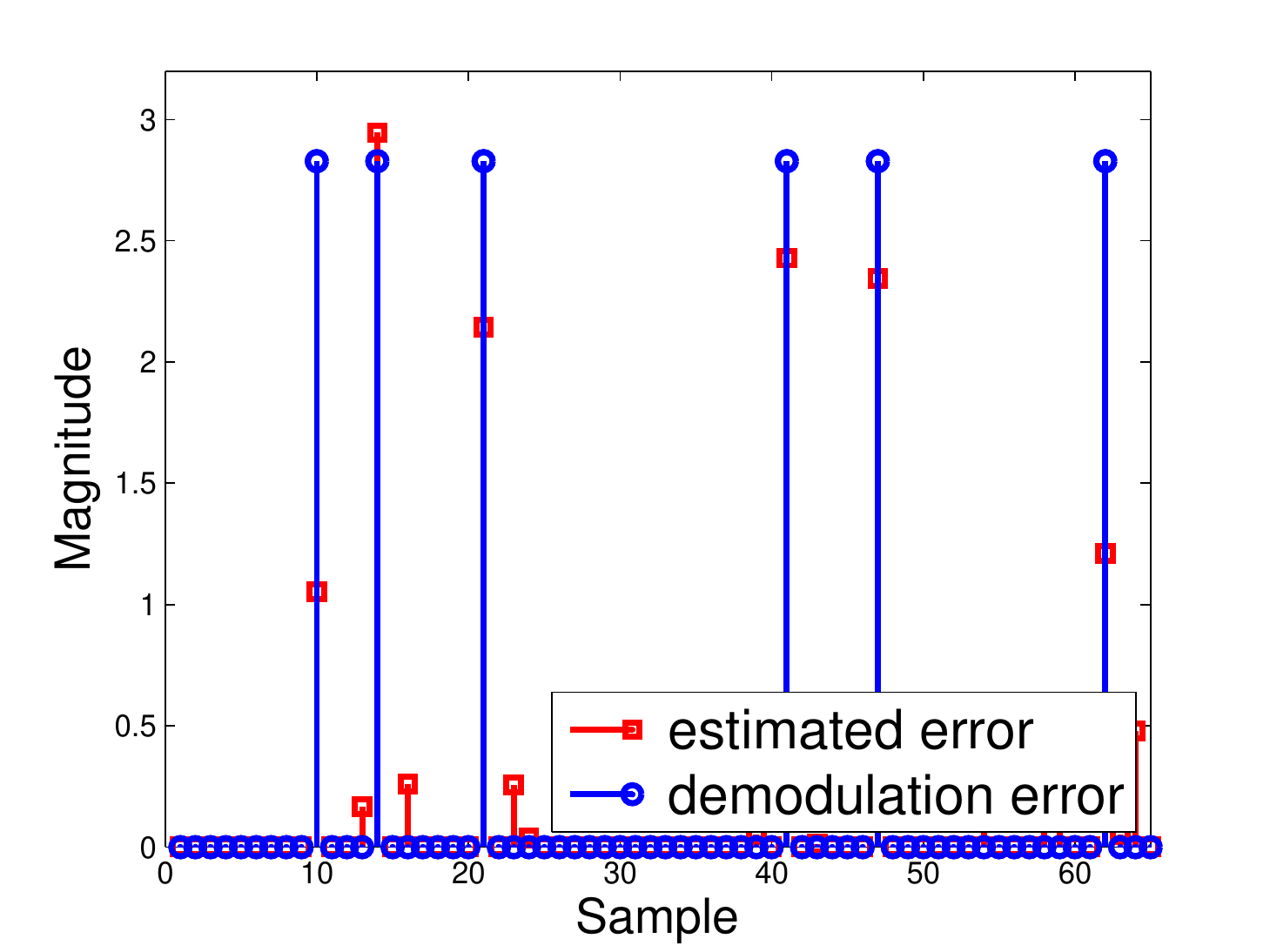}}
	\subfloat[][]{\includegraphics[width=1.8in]{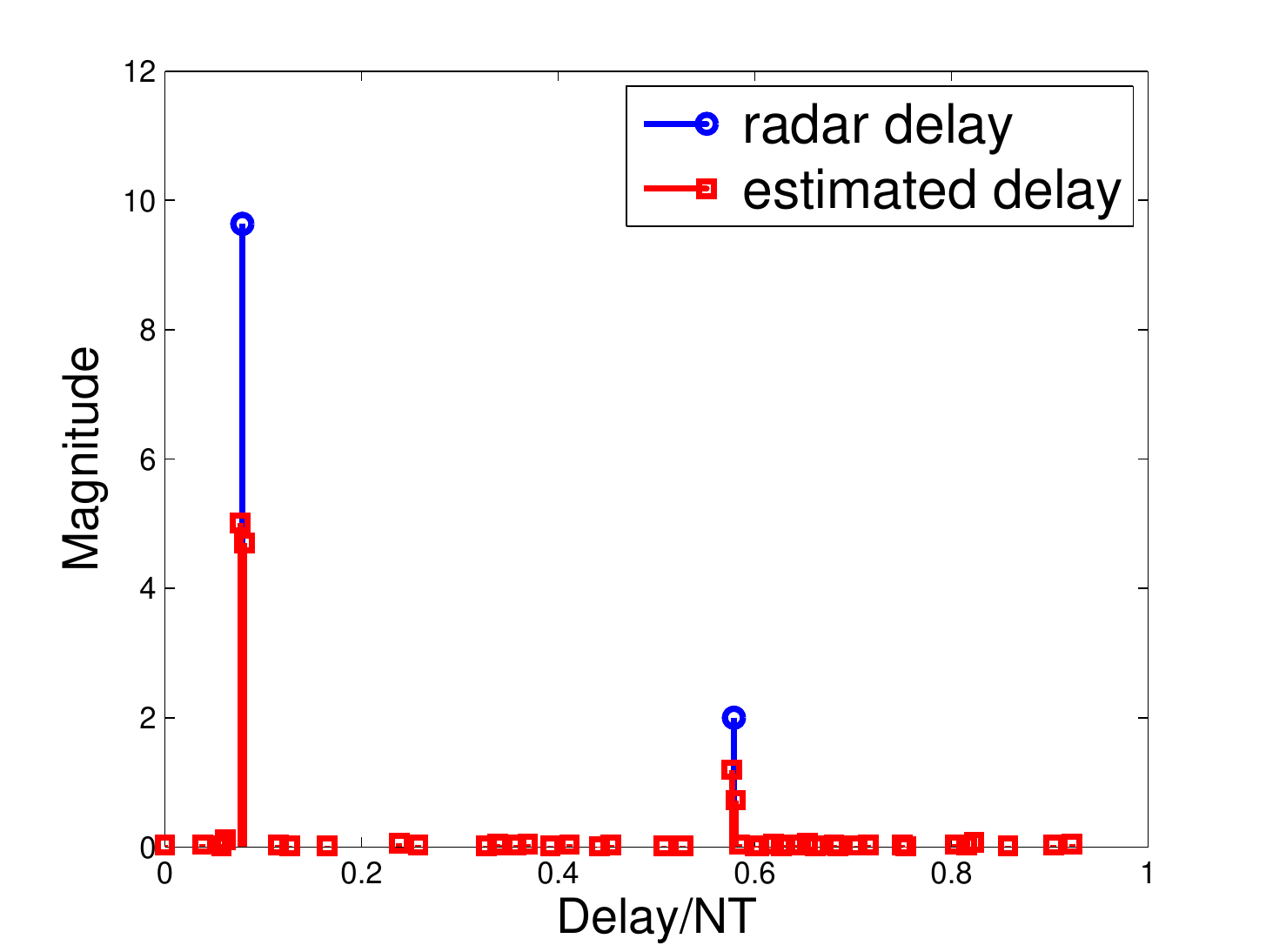}}

	\subfloat[][]{\includegraphics[width=1.8in]{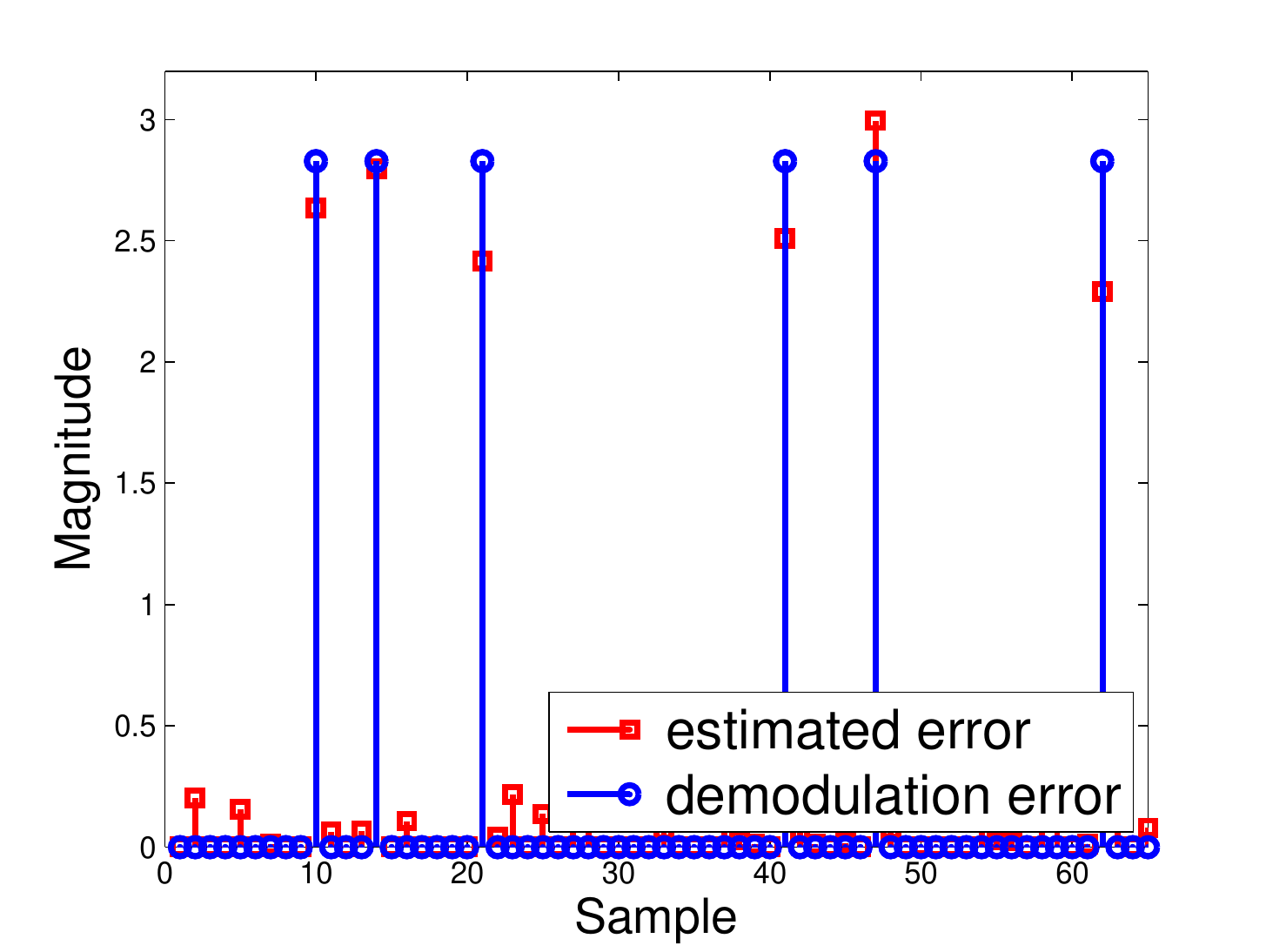}}
	\subfloat[][]{\includegraphics[width=1.8in]{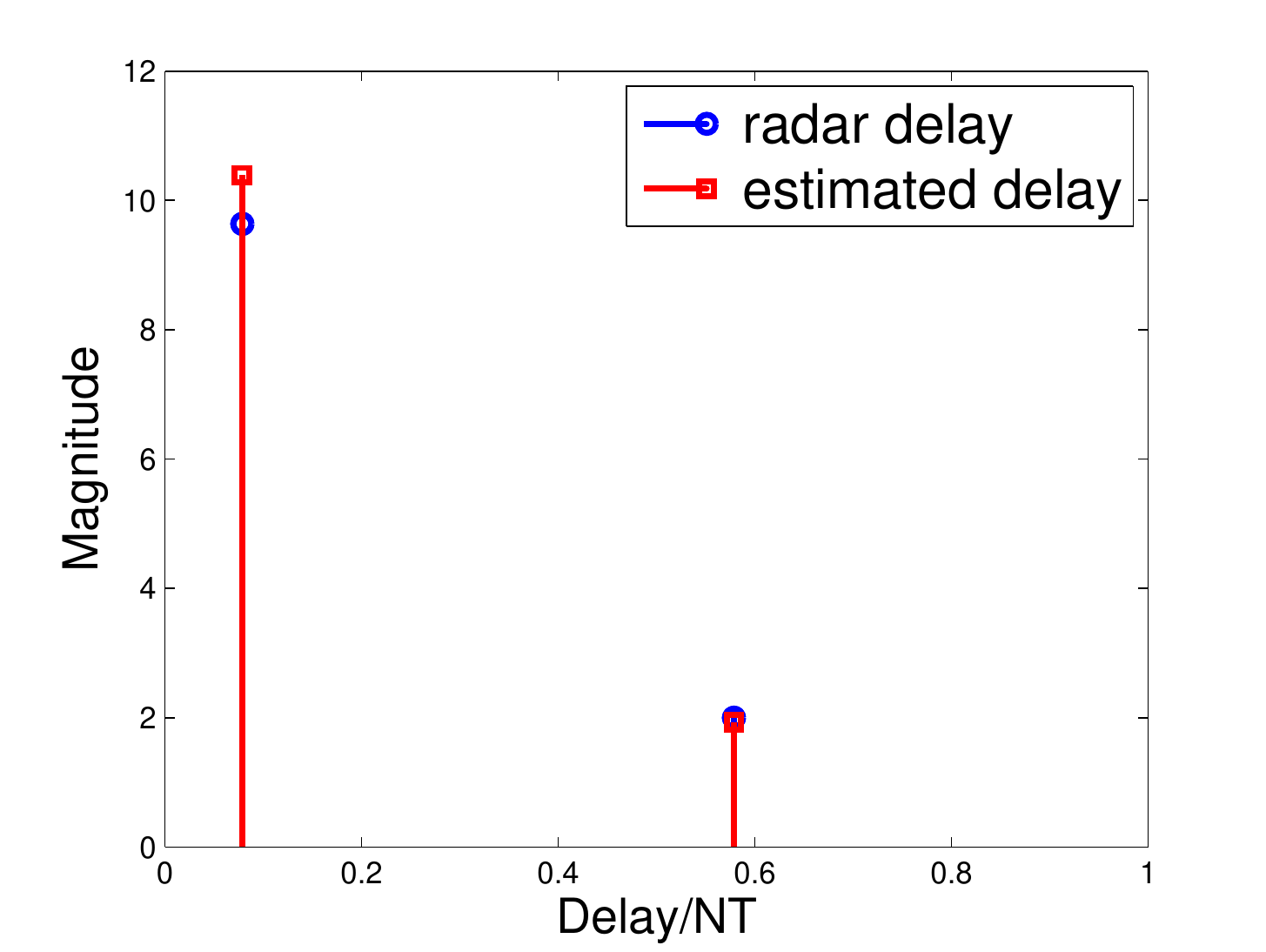}}
	
	\caption{Comparison between (a)(b) CS-L1 algorithm and (c)(d) CS-AN algorithm.}
	\label{fig:example1}
\end{figure}

\begin{figure}
	\centering
	
	\subfloat[][]{\includegraphics[width=1.8in]{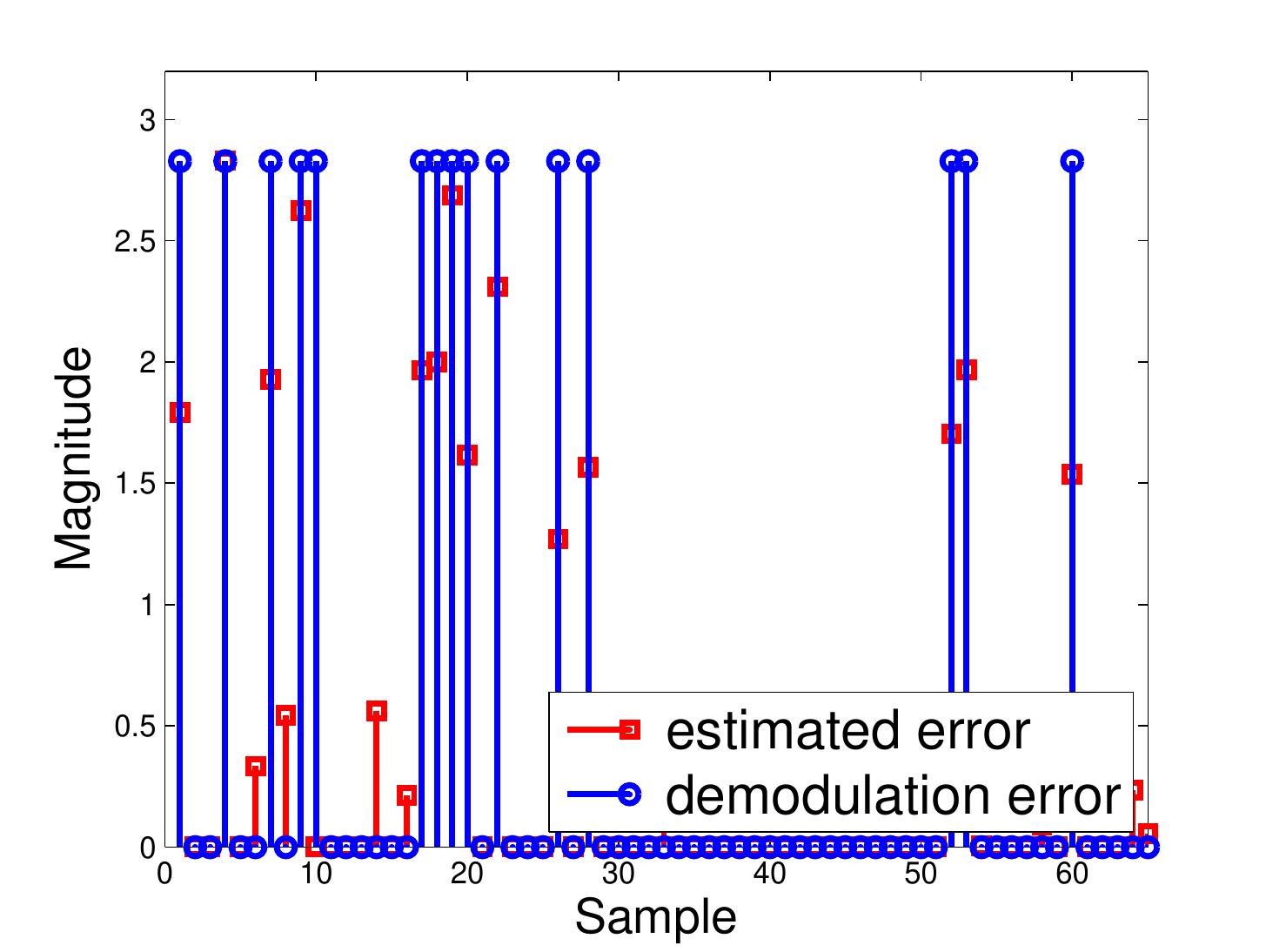}}
	\subfloat[][]{\includegraphics[width=1.8in]{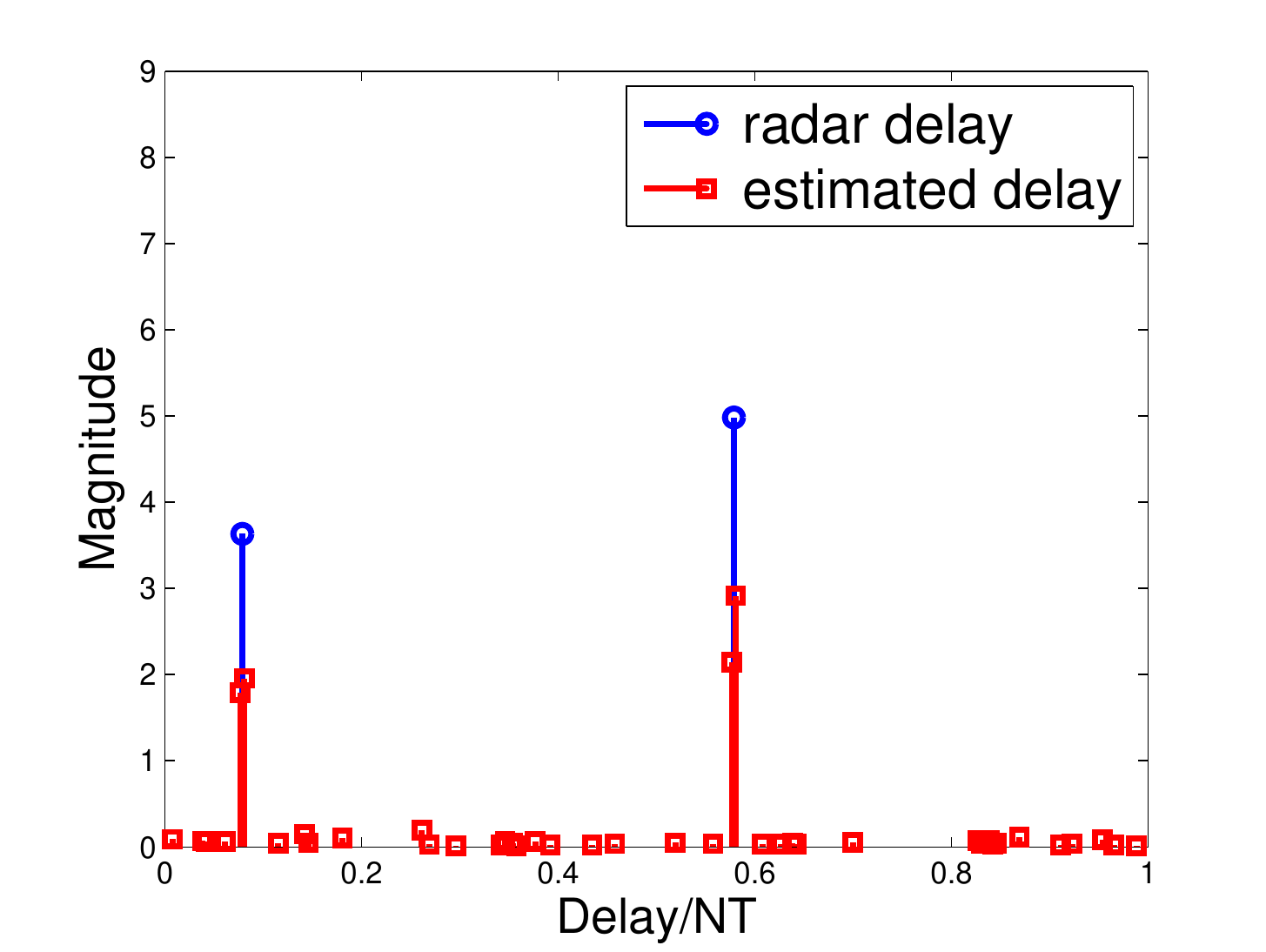}}
	
	\subfloat[][]{\includegraphics[width=1.8in]{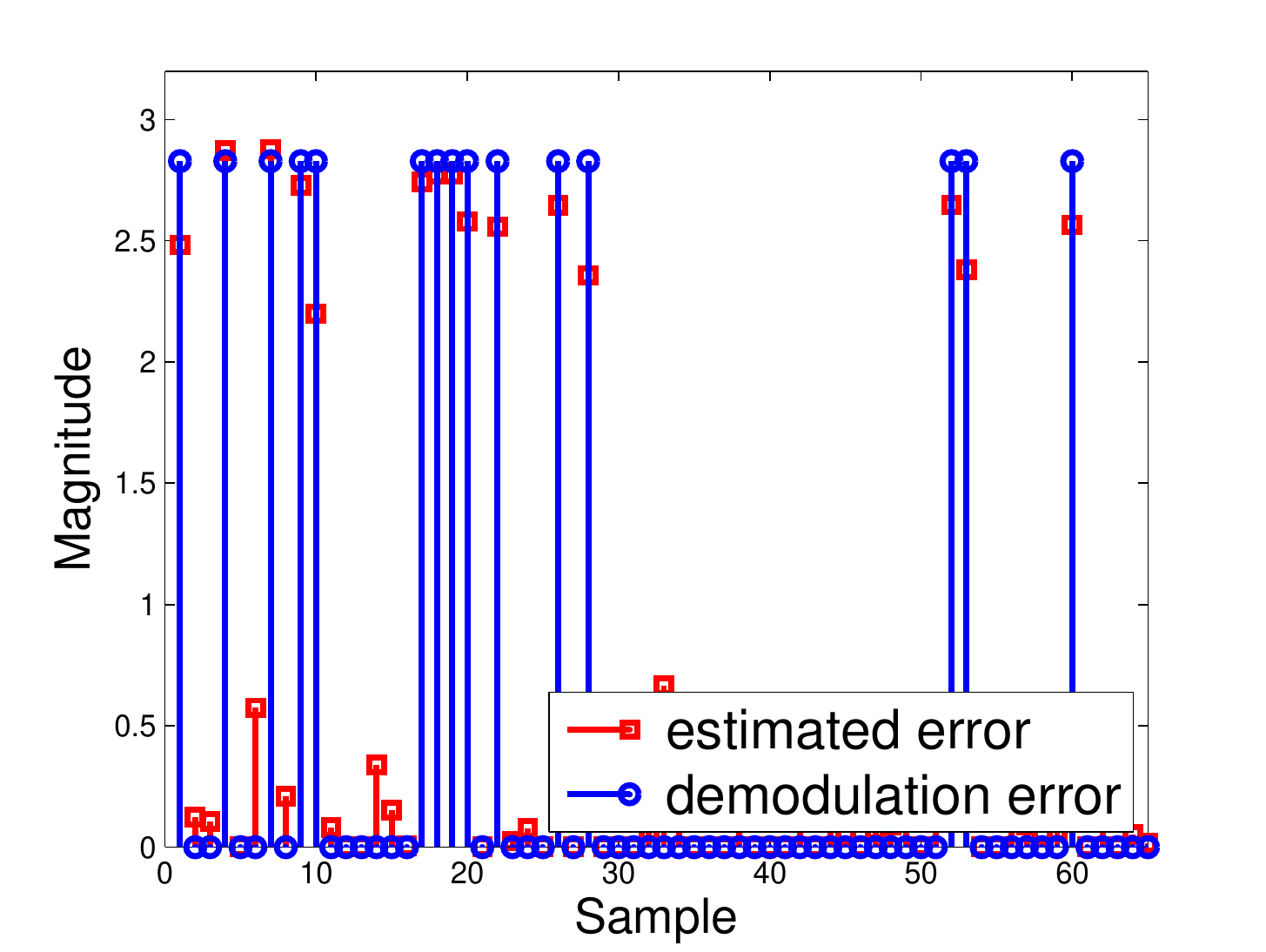}}
	\subfloat[][]{\includegraphics[width=1.8in]{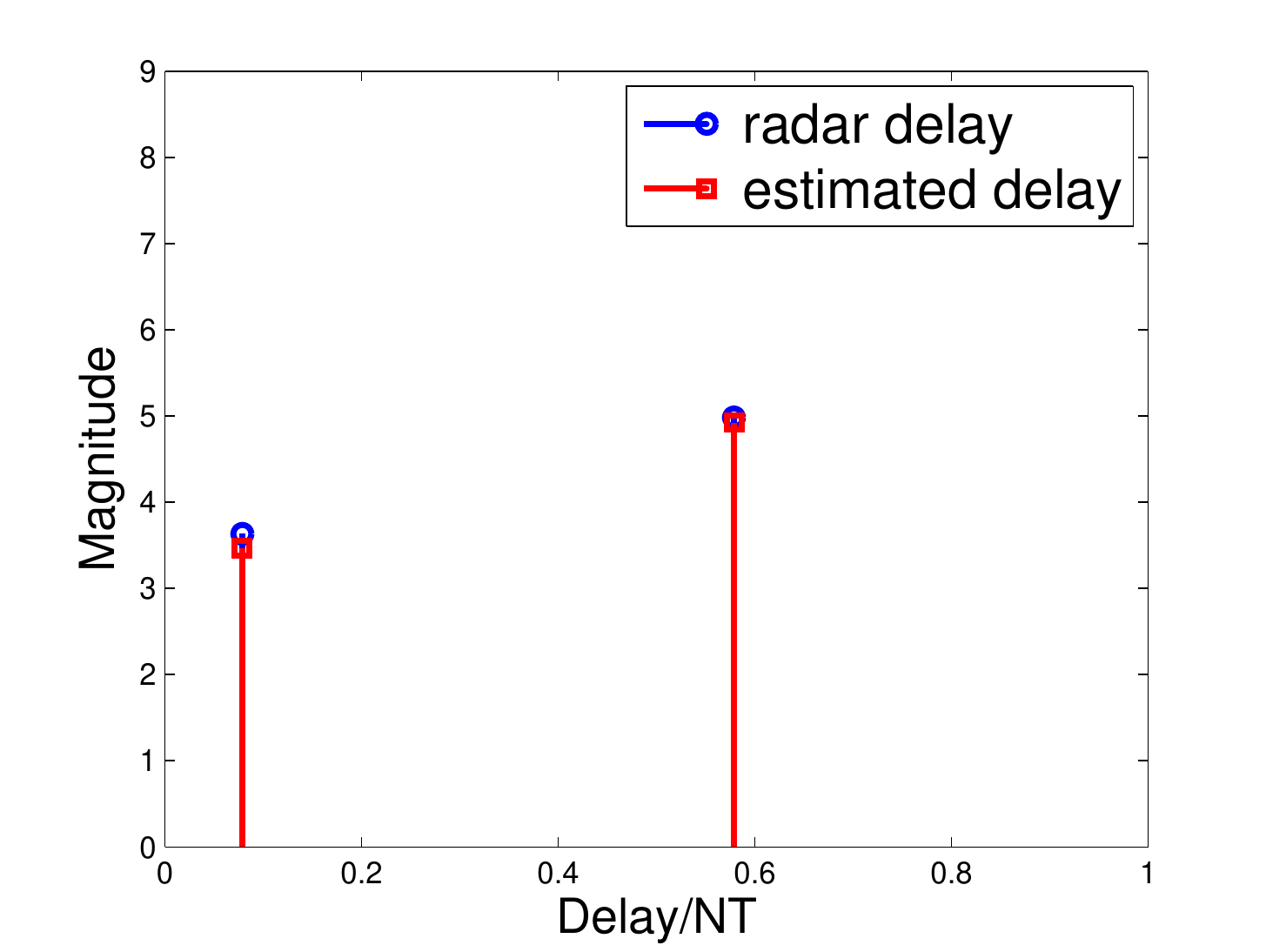}}

	\caption{Comparison between (a)(b) CS-L1 algorithm and (c)(d) CS-AN algorithm.}
	\label{fig:example2}
\end{figure}

Fig. \ref{fig:example1} and Fig. \ref{fig:example2} give the results when the SER is 0.1 and 0.3, respectively. The basis mismatch inherent in the CS-L1 algorithm returns more false alarms already in the initial iteration, but in both cases the number of false detections at $l=0$ is definitely unacceptable. Not surprisingly, the first iteration dramatically cleans the environment, showing that for both algorithms the interference picture becomes much clearer and much closer to the reality: the CS-AN algorithm, however, is definitely superior to the CS-L1 algorithm under both considered values of the SER, which confirms the importance of a proper basis matching in the interference identification-estimation phase.

\section{Fast Algorithm Based on Non-convex Factorization}

As outlined in the previous section, the off-the-shelf solvers for \eqref{eq:atomicnorm} tend to be slow, especially for large-dimensional cases. A possible alternative, in these circumstances, could be the Alternating Direction Method of Multipliers (ADMM) \cite{boyd2011distributed}, which requires an eigenvalue decomposition at each iteration, still entailing a computational complexity ${\cal O}((N+K)^3)$ per iteration, again posing a complexity issue for large-scale problems.

In this section, we derive a fast method for solving this SDP via the non-convex factorization proposed by Burer and Monteiro \cite{burer2003nonlinear}. For notational simplicity, the superscript $l$ of the variables $\bm z$ and $\bm v$ are omitted in what follows. Defining
\begin{eqnarray}
\label{eq:Z}
\bm Z &=& \left[ {\begin{array}{*{20}{c}}
	\bm U & \bm{X}^H\\
	\bm{X}& \bm T
	\end{array}} \right] \in \mathbb{C}^{(N+K) \times (N+K)},
\end{eqnarray}
we rewrite \eqref{eq:SDP} as
\begin{eqnarray}
\label{eq:P1}
\mathop{\min}\limits_{\bm Z,\bm v} && \sum_{k=-\tilde N_1}^{\tilde N_2} \frac{1}{2} {\left| z(k) - \left\langle \bm f_k , \bm H \bm A \bm v \right\rangle - \left\langle \bm X , \bm{\bar d}_k \bm e_k^H \right\rangle \right|^{2}} \nonumber \\
&&+ \frac{\lambda}{2N} {\text{Tr}}\left( \bm U \right) + \frac{\lambda \text{Tr}(\bm T)}{2} + \gamma \|\bm{v}\|_1, \\
\text{s.t.} && {\cal P}_{\rm Toep}(\bm U) = \bm U, \bm Z \succeq 0, \nonumber
\end{eqnarray}
where ${\cal P}_{\rm Toep}(\cdot)$ denotes the projection of the input matrix onto a Toeplitz matrix. In particular, we have
\begin{eqnarray}
{\cal P}_\text{Toep}(\bm{U})&=& \text{Toep}({\cal G}(\bm{U}))
\end{eqnarray}
with ${\cal G}(\bm{U})$ outputing an $N$-dimensional vector whose $(k+1)$-th element is the mean value of the $k$-th subdiagnal elements of the input matrix, i.e., ${\cal G}(\bm{U})_{k+1} = \frac{1}{N-k} \sum_{j=1}^{N-k} U(k+j,j)$.

As will be shown later, the algorithm can be accelerated if the solution to \eqref{eq:P1} is of low-rank. Note that the number of active radars is usually small in practice, which leads to small $\hat J$. We show in the following lemma that under some conditions the rank of the solution for $\bm Z$ equals $\hat J$, which enables us to accelerate the algorithm through non-convex factorization. The proof is given in Appendix B.
\begin{lemma}
	Suppose $\bm{\hat X} = \sum_{j=1}^{\hat J} \hat c_j \bm{\hat h}_j \bm a(\hat \tau_j')^H$ is the solution to \eqref{eq:atomicnorm}. If $N \geq 257$ \footnote{The condition $N \geq 257$ is a technical requirement that originally comes from Theorem 1.3 of \cite{candes2013super}. It is found via simulations that the result still holds without such condition \cite{yang2014exact}.} and $\Delta_{\hat \tau'} \geq \frac{4}{N-1}$ where
	\begin{eqnarray}
	\Delta_{\hat \tau'} = \inf_{\hat \tau_1', \hat \tau_2' \in [0,1]: \hat \tau_1' \neq \hat \tau_2'} \min\left\{ | \hat \tau_1' - \hat \tau_2'|, 1-|\hat \tau_1' - \hat \tau_2'| \right\},
	\end{eqnarray}
	then there exists $\bm{\hat Z}$ as a solution to \eqref{eq:P1} that satisfies $\text{rank}(\bm{\hat Z}) = \hat J$.
\end{lemma}

\vspace{0.2cm}

If an upper bound on the number of active radars, say $\bar J$, is known in advance, then we can introduce the extra constraint $\text{rank}(\bm Z) \leq \bar J$ in \eqref{eq:P1}, whereby restricting the search space to matrices of rank at most $\bar J$ does not change the globally optimal value\footnote{If the upper bound is larger than the true number, it does not affect the performance of the algorithm, but an additional computational cost is incurred because the algorithm searches in a space than the one the solution belongs to. If the upper bound is smaller than the true number, some of the radars can go un-detected and the radar interference cannot be canceled completely. As a result, the demodulation performance decreases as well.}. Additionally, we relax the constraint ${\cal P}_{\rm Toep}(\bm U) = \bm U$, replacing it with the penalty term $\frac{\varrho}{2} \| {\cal P}_{\rm Toep}(\bm U) - \bm U \|_F^2$, so that the problem is recast as
\begin{eqnarray}
\label{eq:P3}
\mathop {\min }\limits_{\bm Z, \bm v} && \zeta(\bm Z, \bm v) =  \frac{\lambda}{2N} {\text{Tr}}\left( \bm U \right) \nonumber \\
&& + \sum_{k=-\tilde N_1}^{\tilde N_2} \frac{1}{2} {\left| z(k) - \left\langle \bm f_k , \bm H \bm A \bm v \right\rangle - \left\langle \bm X , \bm{\bar d}_k \bm e_k^H \right\rangle \right|^{2}} \nonumber \\
&& + \frac{\lambda \text{Tr}(\bm T)}{2} + \frac{\varrho}{2} \| {\cal P}_{\rm Toep}(\bm U) - \bm U \|_F^2 + \gamma \|\bm{v}\|_1, \\
\text{s.t.} && \bm Z \succeq 0, \text{rank}(\bm Z) \leq \bar J. \nonumber
\end{eqnarray}
Setting $\bm Z = \bm V \bm V^H$ where $\bm V \in \mathbb{C}^{(N+K) \times \bar J}$, \eqref{eq:P3} becomes an unconstrained optimization of $\min_{\bm V, \bm v} \zeta(\bm V \bm V^H, \bm v)$.  Though this unconstrained problem is non-convex, its dimension is lower than that of the original problem in \eqref{eq:P1} and has no conic constraint, which leads to reduced computational complexity. A very effective means to undertake the desired minimization of $\zeta(\bm V \bm V^H, \bm v)$ is to resort to a Conjugate Gradient (CG) algorithm \cite{dai2013nonlinear}, which is a fast first-order algorithm. The algorithm requires the objective function to be smooth, therefore we approximate $\| \cdot \|_1$ in $\zeta(\bm V \bm V^H, \bm v)$ with the convex, differentiable function:
\begin{eqnarray}
\| \bm v \|_1 &\approx& \psi_\mu (\bm v) \nonumber \\
&=& \mu \sum_{m=0}^{M-1} \log \left( \frac{\exp(|v_m|/\mu)+\exp(-|v_m|/\mu)}{2} \right) \nonumber \\
&=& \mu \sum_{m=0}^{M-1}\log \cosh(|v_m|/\mu)
\end{eqnarray}
where $\mu$ controls the smoothing level
\footnote{A similar approach has also been used in \cite{sun2017complete} for the smoothing of the $\ell_1$-norm in the objective function. In fact, there is nothing special about this choice and we believe that some other twice continuously differentiable approximation to $\| \cdot \|_1$ would work and yield qualitatively similar results.}. Hence, the problem becomes
\begin{eqnarray}
	\label{eq:P5}
	\mathop {\min }\limits_{\bm V, \bm v} && \tilde \zeta(\bm V \bm V^H, \bm v),
\end{eqnarray}
where
\begin{eqnarray}
\label{eq:tildezeta}
\tilde \zeta(\bm Z, \bm v) &=& \sum_{k=-\tilde N_1}^{\tilde N_2} \frac{1}{2} {\left| z(k) - \left\langle \bm f_k , \bm H \bm A \bm v \right\rangle - \left\langle \bm X , \bm d_k \bm e_k^H \right\rangle \right|^{2}} \nonumber \\
&& + \frac{\lambda}{2N} {\text{Tr}}\left( \bm U \right) + \frac{\lambda \text{Tr}(\bm T)}{2} \nonumber \\
&&  + \frac{\varrho}{2} \| {\cal P}_{\rm Toep}(\bm U) - \bm U \|_F^2 + \gamma \psi_\mu (\bm{v}).
\end{eqnarray}

The minimization problem \eqref{eq:P5} can be effectively solved using the CG algorithm, undertaking the iteration
\begin{eqnarray}
\label{eq:updateV}
\bm V^{(k)} &=& \bm V^{(k-1)} + \varsigma_k \bm P_k, \\
\label{eq:updatev}
\bm v^{(k)} &=& \bm v^{(k-1)} + \varsigma_k \bm p_k,
\end{eqnarray}
where $\varsigma_k$ is the step size, $\bm p_k$ and $\bm P_k$ are the search directions at step $k$, evaluated as the weighted sums of the gradients at present iteration and the search direction used at the previous one. Specifically, if $\nabla_{\bm V} \tilde \zeta( \bm V^{(k-1)} (\bm V^{(k-1)})^H, \bm v^{(k-1)})$ and $\nabla_{\bm v} \tilde \zeta( \bm V^{(k-1)} (\bm V^{(k-1)})^H, \bm v^{(k-1)})$ denote the gradients of $\tilde \zeta( \bm V \bm V^H, \bm v)$ at the $k$-th iteration, then we have
\begin{eqnarray}
\label{eq:P}
\bm P_k &=& -\nabla_{\bm V} \tilde \zeta( \bm V^{(k-1)} (\bm V^{(k-1)})^H, \bm v^{(k-1)}) + \omega_k \bm P_{k-1}, \\
\label{eq:p}
\bm p_k &=& -\nabla_{\bm v} \tilde \zeta( \bm V^{(k-1)} (\bm V^{(k-1)})^H, \bm v^{(k-1)}) + \omega_k \bm p_{k-1},
\end{eqnarray}
where
\begin{eqnarray}
\label{eq:HS}
\omega_k = \frac{\left[ \begin{array}{l}
	\left\langle \nabla_{\bm V} \tilde \zeta( \bm V^{(k-1)} (\bm V^{(k-1)})^H, \bm v^{(k-1)}) , \bm Q_{k-1}  \right\rangle \\
	+ \left\langle \nabla_{\bm v} \tilde \zeta( \bm V^{(k-1)} (\bm V^{(k-1)})^H, \bm v^{(k-1)}) , \bm q_{k-1}  \right\rangle
	\end{array} \right] }{ \left\langle \bm P_{k-1} , \bm Q_{k-1} \right\rangle + \left\langle \bm p_{k-1} , \bm q_{k-1} \right\rangle},
\end{eqnarray}
with $\bm Q_{k} = \nabla_{\bm V} \tilde \zeta( \bm V^{(k)} (\bm V^{(k)})^H, \bm v^{(k)}) - \nabla_{\bm V} \tilde \zeta( \bm V^{(k-1)} (\bm V^{(k-1)})^H, \bm v^{(k-1)})$ and $\bm q_{k} = \nabla_{\bm v} \tilde \zeta( \bm V^{(k)} (\bm V^{(k)})^H, \bm v^{(k)}) - \nabla_{\bm v} \tilde \zeta( \bm V^{(k-1)} (\bm V^{(k-1)})^H, \bm v^{(k-1)})$. The expressions of the gradients are derived in Appendix C.

For clarity, we summarize the proposed non-convex solver in Algorithm 2. To guarantee that the objective function does not increase with $k$, the CG  utilizes the Armijo line search \cite{boumal2015low} (line 6 of Algorithm 2), so that the algorithm converges to a stationary point of the surrogate problem, namely, the point where the smoothed objective function \eqref{eq:tildezeta} has vanishing gradient.

\begin{algorithm}[htbp]
	\label{tab:A2}
	\caption{Conjugate gradient algorithm}
	\begin{tabular}{lcl}
		Input $\bar{\bm D}$, $\bm z$, $N$, $\lambda$, $\gamma$, $\epsilon$, $\varrho$, $\mu$.\\
		1, Initialize $\bm V^{(0)}$, $\bm v^{(0)}$, $k=0$. \\
		\sf{Do} \\
		2,\hspace{0.4cm}  $k = k+1$.\\
		3,\hspace{0.4cm}  Compute $\nabla_{\bm V} \tilde \zeta( \bm V^{(k-1)} (\bm V^{(k-1)})^H, \bm v^{(k-1)})$ and \\
		\hspace{0.6cm} $\nabla_{\bm v} \tilde \zeta( \bm V^{(k-1)} (\bm V^{(k-1)})^H, \bm v^{(k-1)} )$. \\
		\hspace{0.4cm} \sf{If $k = 1$} \\
		4,\hspace{0.8cm}  $\bm P_k = -\nabla_{\bm V} \tilde \zeta( \bm V^{(k-1)} (\bm V^{(k-1)})^H, \bm v^{(k-1)})$, \\
		\hspace{1cm} $\bm p_k = -\nabla_{\bm v} \tilde \zeta( \bm V^{(k-1)} (\bm V^{(k-1)})^H, \bm v^{(k-1)})$, \\
		\hspace{0.4cm} \sf{Else } \\
		5,\hspace{0.8cm}  Calculate $\bm P_k$ and $\bm p_k$ according to \eqref{eq:P} and\\
		\hspace{1cm} \eqref{eq:p} where $\omega_k$ is obtained by \eqref{eq:HS}. \\
		\hspace{0.4cm} \sf{End if} \\
		6,\hspace{0.4cm}  Update $\bm V^{(k)}$ and $\bm v^{(k)}$ according to \eqref{eq:updateV} and \eqref{eq:updatev} \\
		\hspace{0.6cm} where $\varsigma_k$ is obtained via Armijo line search. \\
		\sf{While $\| \nabla_{\bm V} \tilde \zeta( \bm V^{(k)} (\bm V^{(k)})^H) \|_2 \leq \epsilon$.} \\
		7, Obtain $\bm{\hat X}$ according to \eqref{eq:Z} with $\bm{\hat Z} = \bm V^{(k)} (\bm V^{(k)})^H$. \\
		\midrule
		Return $\bm{\hat v} = \bm{v}^{(k)}$, $\bm{\hat X}$.\\
	\end{tabular}
\end{algorithm}

\begin{figure}
	\centering
	
	\includegraphics[width=3.2in]{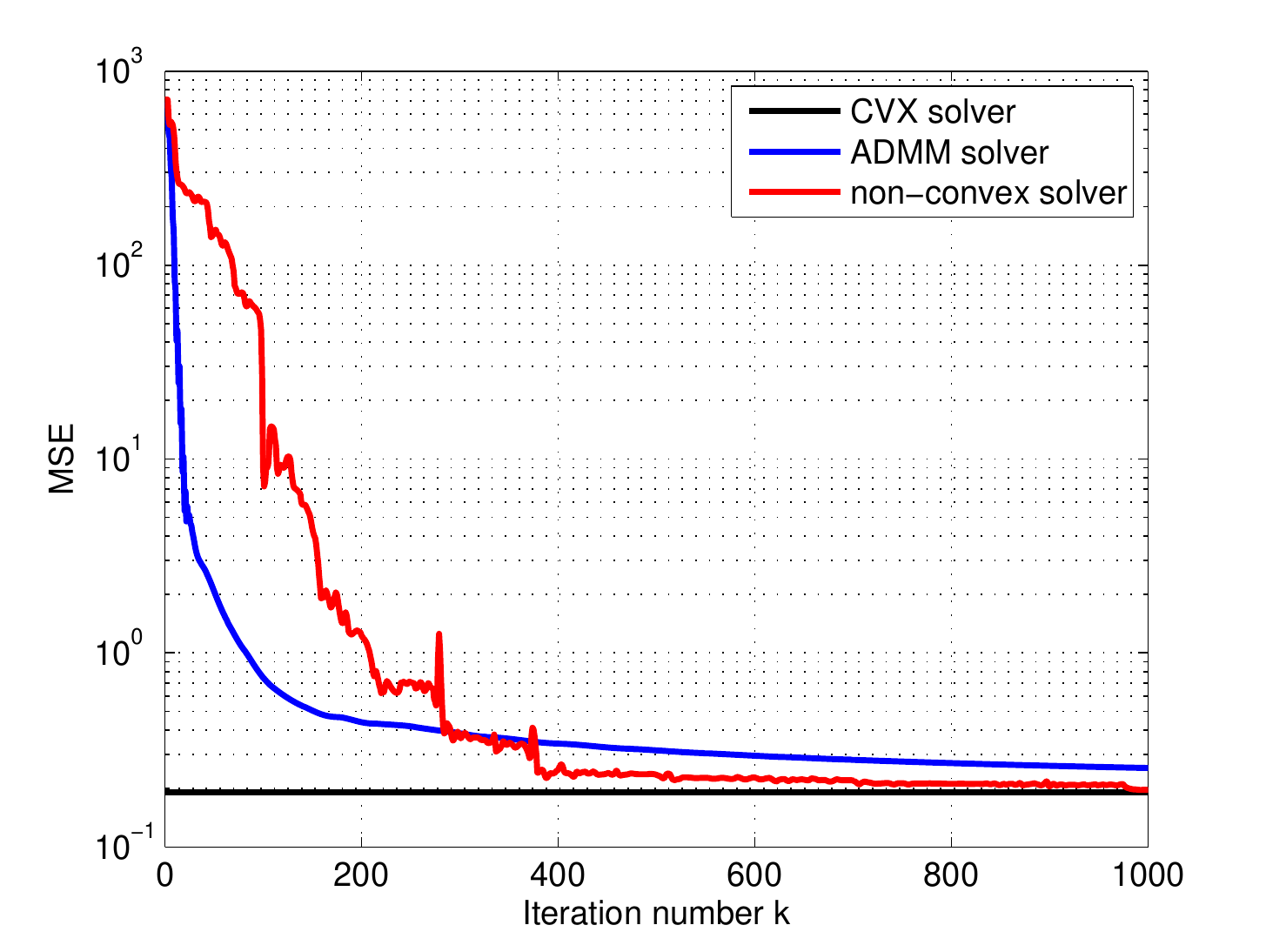}
	
	\caption{Convergence behavior of the proposed non-convex solver. The non-convex solver takes 41.2 seconds with 1000 iterations. The ADMM solver takes 104.2 seconds with 1000 iterations. The CVX solver takes 228.7 seconds. The experiments were carried out on an Intel Xeon desktop computer with a 3.5 GHz CPU and 24 GB of RAM.}
	\label{fig:speed}
\end{figure}

The computational complexity of the proposed algorithm at each iteration is mainly determined by the calculation of $\bm V \bm V^H$, whose complexity is ${\cal O}((N+K)^2 \bar J)$. As $\bar J$ is much smaller than $(N+K)$, the complexity per-iteration is much smaller than that of a classical eigenvalue decomposition, whereby, for large-dimensional problems, the proposed non-convex approach can be faster than those based on the first-order methods such as ADMM and projected gradient descent. We illustrate this fact through a simulation example, whose results are reported in Fig. \ref{fig:speed}. The dimension of the signal is $N=257$, while the other parameters are $J = 2$, $L = 5$ and $K = 5$. The variance of the noise is $\sigma_w^2=0.01$. The non-convex solver is implemented by solving \eqref{eq:P5} with the CG algorithm. We compare the MSE of the proposed algorithm with that given by solving \eqref{eq:SDP} with CVX \cite{cvx} and ADMM solver. As can be seen from Fig. \ref{fig:speed}, the result of the proposed algorithm is close to the solution given by the CVX after 1000 iterations: since the proposed algorithm is also much faster than the ADMM, it appears much more suitable for real-time implementation.

\section{Simulation Results}
\subsection{Simulation Setup}
In order to demonstrate the performance of the proposed algorithms, we simulate a scenario with multiple radars and one communication receiver. The communication system uses OFDM signal with frequency spacing between adjacent subcarrier of 10kHz, $N = 129$ subcarrier frequencies and total bandwidth 1.29MHz. The symbol length is $T = 100 \mu s$ and Quadrature Phase-Shift Keying (QPSK) modulation is used. The channel matrix is generated as  $\bm F^H \bm \Gamma \bm F$ where $\bm \Gamma$ is an identity matrix whose diagonal elements are complex random variables whose moduli follow a Ricean distribution: as a consequence, we model the ray impinging on the receiver as the superposition of a non-fading component, with power $\rho^2$ say, and a fluctuating (zero-mean) component, with mean square value $\sigma^2_h$, so that the Rician factor is $\frac{\rho^2}{\sigma^2_h}$ \cite[p. 79]{goldsmith}.
As to the active radars, they are modeled as point sources in our simulations. For simplicity, the amplitudes of the path gains $c_j$ are generated with fixed magnitude and random phase, and the magnitude is controlled by the power of the paths. We define the signal-to-interference ratio (SIR) at the communication RX as the power ratio of the communication signal and the radar interference. Specifically, the SNR and SIR are defined as
\begin{eqnarray}
\text{SNR} &=& \frac{(\rho^2 + \sigma_h^2)\varepsilon_b}{\sigma_w^2},\\
\text{SIR} &=& \frac{N \varepsilon_b(\rho^2 + \sigma_h^2)}{\left\| \sum_{j=1}^{J} c_j \bm{g}_j \right\|_2^2},
\end{eqnarray}
where $\varepsilon_b$ is the average energy of the signal constellation. Notice that, for fixed SNR, lower values of $\rho$ account for larger fluctuations, up to the limit $\rho=0$, $\text{SNR}=\frac{\sigma_h^2\varepsilon_b}{\sigma_w^2}$, which corresponds to a Rayleigh-fading channel. In what follows we use $\frac{\rho}{\sigma_h} = 3$, unless otherwise specified and then study the impact of the channel fluctuation on the achievable performance.

\begin{figure*}
	\centering
	
	\subfloat[][]{\includegraphics[width=3in]{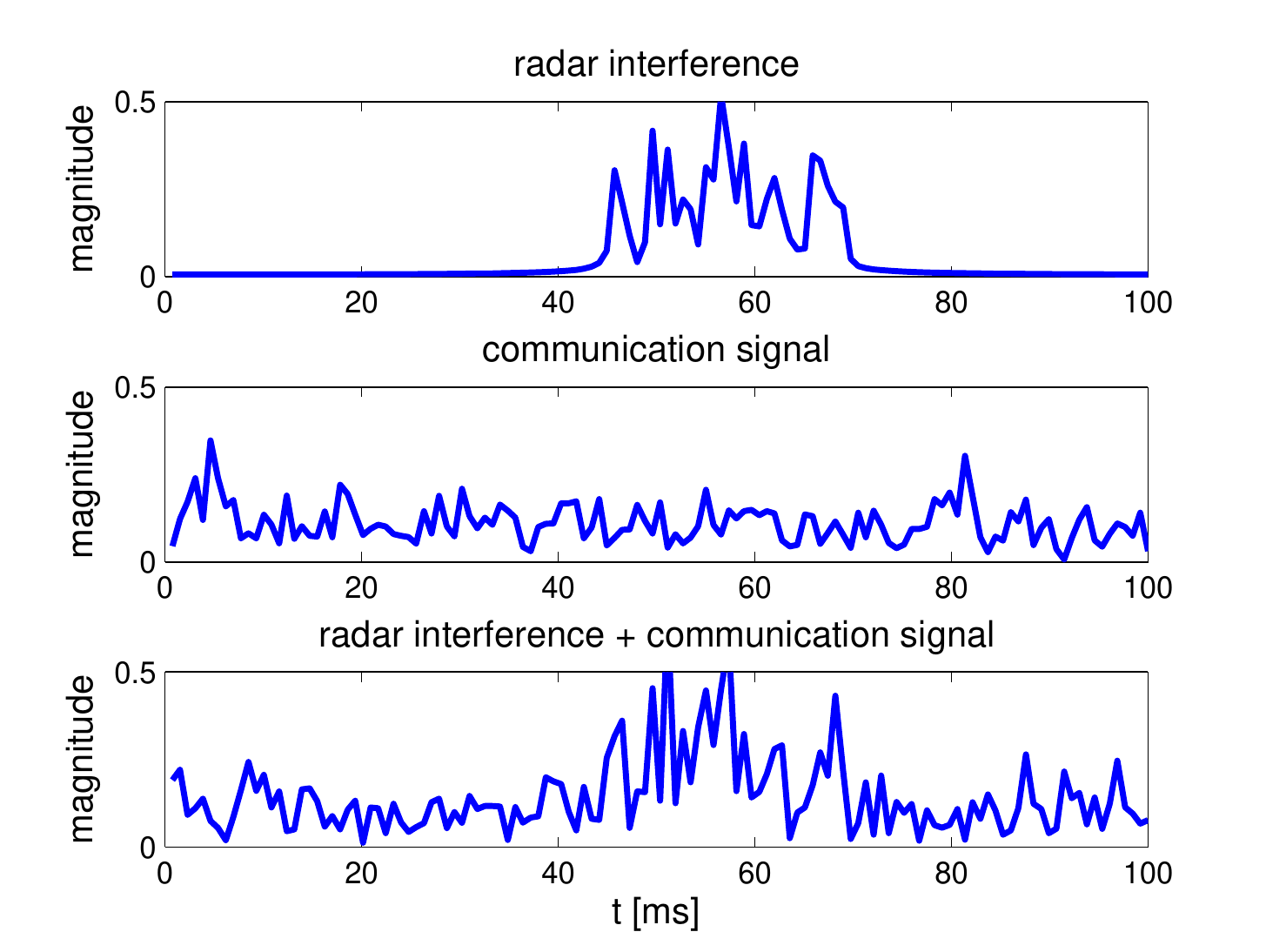}}
	\subfloat[][]{\includegraphics[width=3in]{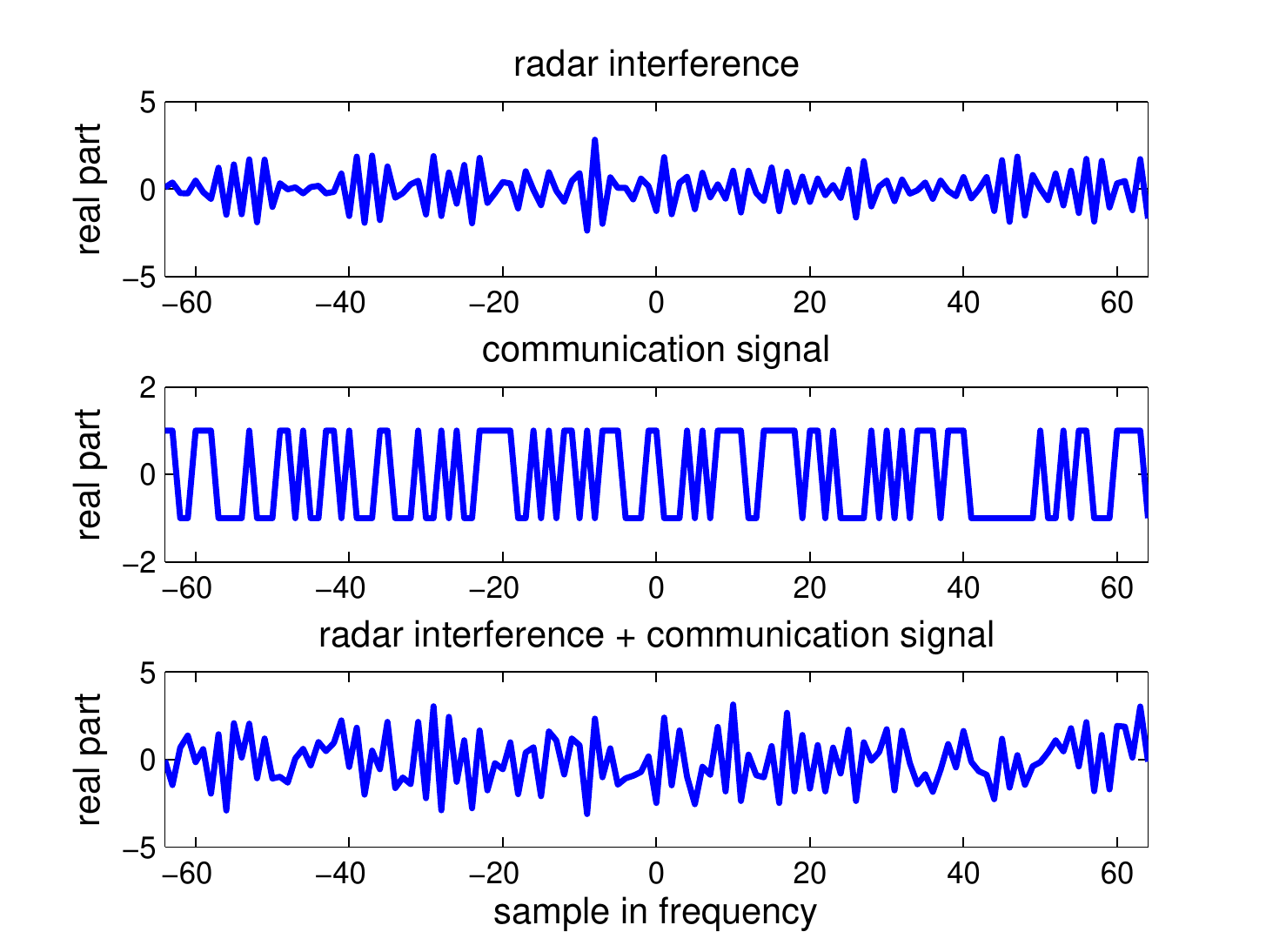}}
	
	\caption{Plots of radar interfernce, communication signal and the received signal of communication RX in (a) time domain and (b) frequency domain. In (a), the magnitude of the signal is plotted against time. In (b), the real part of the signal is plotted versus frequency sample.}
	\label{fig:example}
\end{figure*}

In keeping with the model of Section II, $\bm{\bar g}_j$ lives in a low-dimensional subspace spanned by the columns of $\bar{\bm D}$ matrix, i.e., $\bm{\bar g}_j = \bar{\bm D} \bm h_j$. We use the setting that $\bar{\bm D} = \bm F \bm D$ where $\bm D = [\bm d_1, \bm d_2,..., \bm d_K] \in \mathbb{C}^{N \times K}$ with $\bm d_k \in \mathbb{C}^{N \times 1}$ and $K=5$. In our simulation, the radars use pulse waveforms and each pulse uses Gaussian random code with length $N' = 32$. Specifically, $\bm d_k$ satisfies $d_k(m) \sim {\cal CN}(0, 1/N')$ for $1 \leq m \leq N'$ and $d_k(m) = 0$ for $N' + 1 \leq  m \leq N$. The columns of $\bar{\bm D}$ can be obtained by taking the DFT of $\bm{d}_k$. Fig. \ref{fig:example} gives an example of the signal at the communication RX when $J = 1$. The SIR of the example is set as 0dB: The figure clearly demonstrates how dramatic the effect of even a single co-existing radar can be.

Some other parameters of the simulations are given as follows.

1, The $\tau_j$'s are randomly generated between $10 \mu s$ and $70 \mu s$ for $j=1,2,...,J$.

2, For comparison purposes, we also show the performance of $\bm{\hat b} = \Psi(\bm{\bar r})$, i.e. of a demodulator operating on the raw data: this is named ``Iteration 0", since its result is the initial point to be provided to the iterative algorithms.

3, For the proposed CS-L1 algorithm, we discretize the continuous parameter space to a finite set of grid points of cardinality $\tilde J = 4 N$. The weights for CS-L1 are $\tilde \lambda = \sigma_w \kappa \sqrt{2 \log(\tilde J K)}$ and $\tilde \gamma = 0.5 \sigma_w \sqrt{2 \log(\tilde J K)}$, where $\kappa$ is the average norm of the column vectors of matrix $\bm \Upsilon$.

4, For the proposed CS-AN algorithm, the weights are set as $\lambda = \sigma_w \sqrt{K N \log(K N)}$ and $\gamma = \frac{\lambda}{\sqrt{N}}$. We use the proposed non-convex algorithm to solve \eqref{eq:P5} where $\bar J = 10$, $\mu = 0.01$ and $\varrho = 5$. The algorithm stops as the norm of the gradient is smaller than 0.01.

5, We evaluate the root-mean-squared-error (RMSE) of the radar delay estimation and the relative mean-squared-error (MSE) \footnote{We use the relative MSE rather than the RMSE to evaluate the accuracy because it reflects the loss in energy.} of the estimated waveform  for the proposed CS-L1 and CS-AN algorithms. Notice that the algorithms return a bunch of $\hat \tau_j$'s, which can be either true detections or false alarms, and a radar cannot be identified if there is no $\hat \tau_j$ close to its position. We thus refer to the RMSE conditioned on correct radars identification: in undertaking simulation, a radar is declared to be correctly identified if there is a $\hat \tau_j$ whose error is smaller than $T/4$, which is the grid sizes of the simulated CS-L1 algorithm. Specifically, the delay RMSE and relative waveform MSE are calculated as
\begin{eqnarray}
\text{RMSE}_\tau &=& \sqrt{\frac{1}{\text{MC}} \sum_{m=1}^{\text{MC}}  \frac{1}{|\Omega_m|}\sum_{j \in \Omega_m} (\tau_j^{(m)} - \hat \tau_j^{(m)})^2}, \\
\text{MSE}_{cg} &=& \frac{1}{\text{MC}} \sum_{m=1}^{\text{MC}} \frac{1}{|\Omega_m|} \sum_{j \in \Omega_m} \frac{ \left\| c_j^{(m)} \bm g_j^{(m)} - \hat c_j^{(m)} \bm{\hat g}_j^{(m)} \right\|_2^2 } {\left\|c_j^{(m)} \bm g_j^{(m)} \right\|_2^2} , \nonumber \\
\end{eqnarray}
respectively, where $\text{MC}$ is the number of runs; $\Omega_m$ is the index set of the identified radars in the $m$-th simulation; $|\cdot|$ denotes the cardinality of the input set; $\tau_j^{(m)}$, $c_j^{(m)}$ and $\bm g_j^{(m)}$ are the delay, path gain and waveform of the $j$-th radar in the $m$-th run, respectively, while $\hat \tau_j^{(m)}$ and $\hat c_j^{(m)} \bm{\hat g}_j^{(m)}$ are the respective estimates.

\subsection{Performance}

\begin{figure}
	\centering
	
	\subfloat{\includegraphics[width=3in]{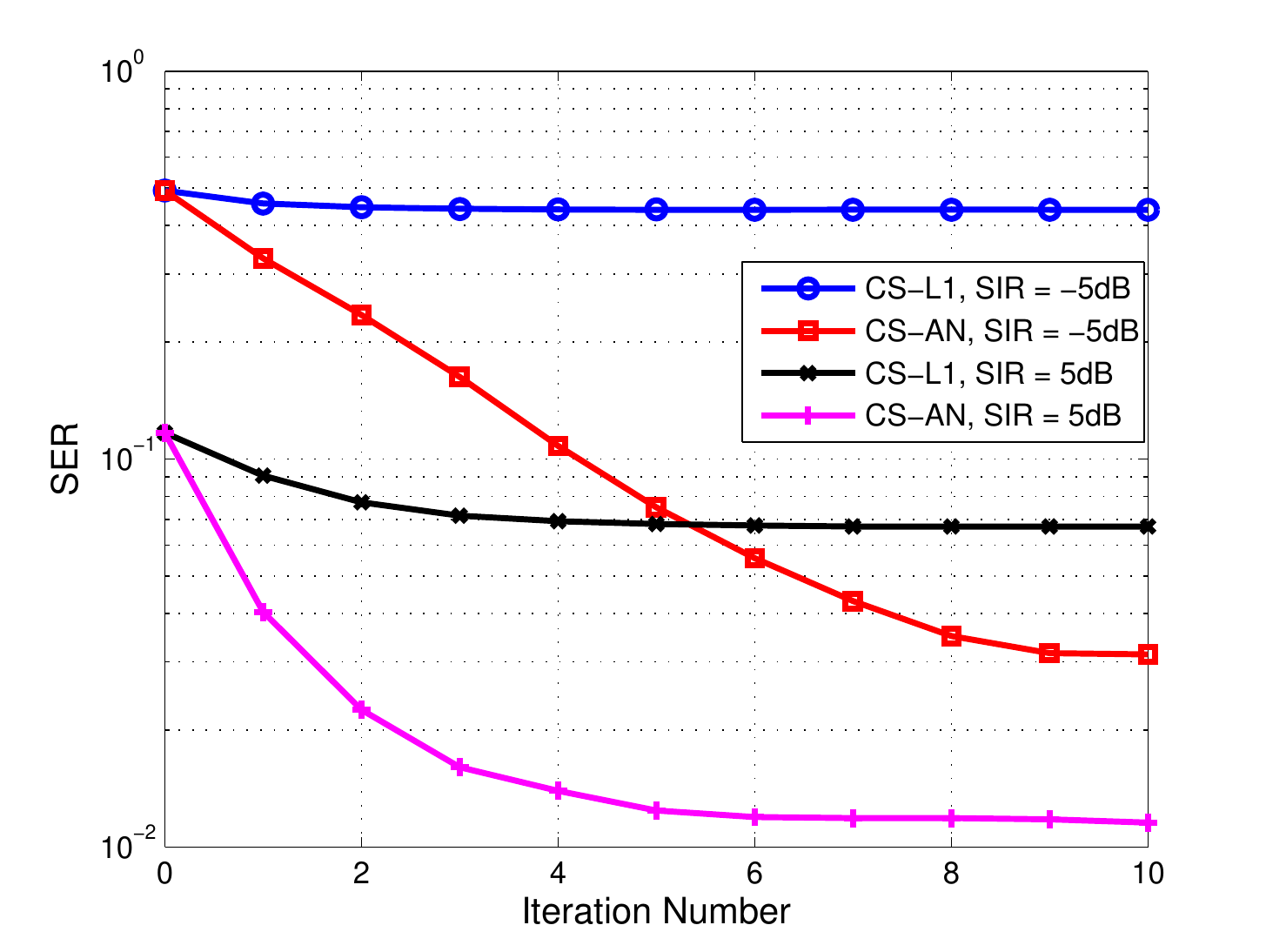}}
	
	\caption{Plots of SER against the iteration number. The SNR is set as 15dB in the simulations.}
	\label{fig:iteration}
\end{figure}

We compare the SER performance of the proposed algorithms. The number of active radars is set as $J=2$ in the simulation. We firstly analyze the impact of iterations on the performance of the proposed CS-L1 and CS-AN algorithms. In Fig. \ref{fig:iteration}, the SER of the algorithms are plotted against the iteration number. As can be seen from the figure, both the CS-L1 and the CS-AN algorithms converge monotonically, but CS-AN converges to a much smaller demodulation error. The algorithms usually converge within 10 iterations.

\begin{figure*}
	\centering
	
	\subfloat[][]{\includegraphics[width=3in]{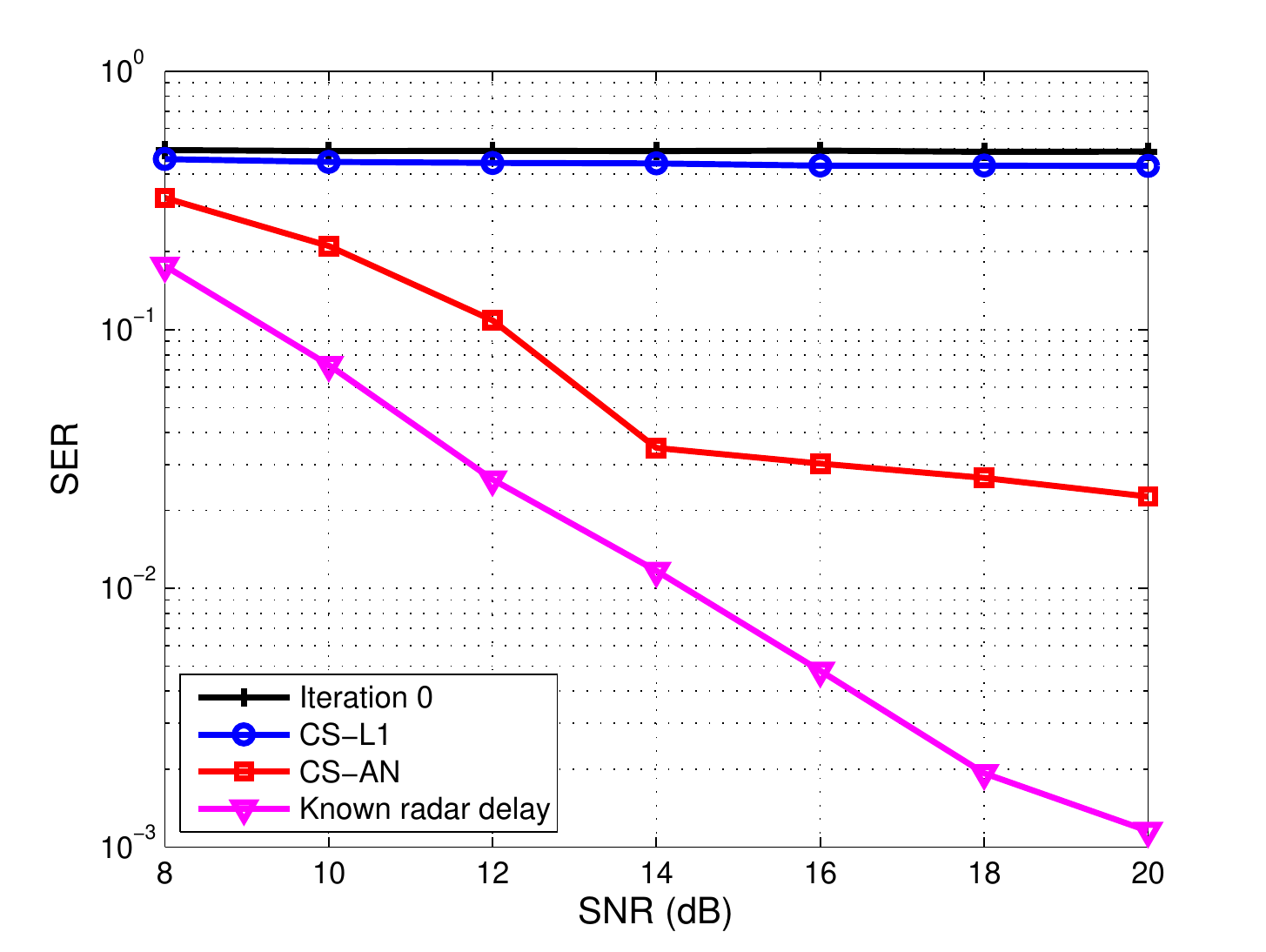}}
	\subfloat[][]{\includegraphics[width=3in]{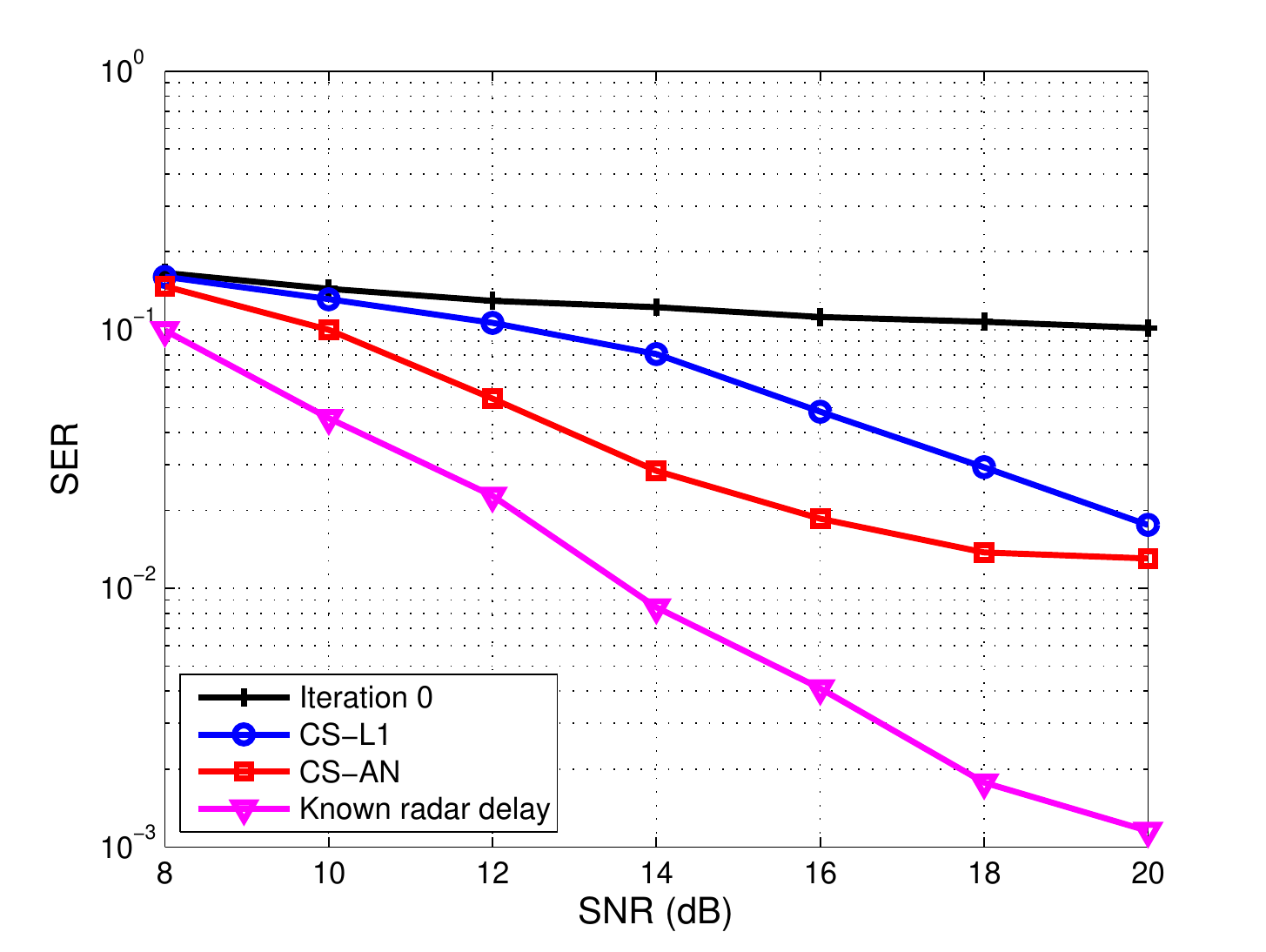}}

	\caption{Comparison of the algorithm SERs when the SIR of communication is (a) -5dB and (b) 5dB. }
	\label{fig:SER}
\end{figure*}

In Fig. \ref{fig:SER}, the effect of the SNR is studied: both proposed CS-L1 and CS-AN algorithms provide better SER performance than Iteration 0. The CS-AN algorithm also outperforms the CS-L1 algorithm in all situations. In order to quantify the cost incurred by the lack of cooperation between the active radars and the communication system, we also report the performance of the algorithm assuming known radar delays, i.e. of the modified CS-L1 algorithm presented in Section III-A: in undertaking the simulation, we set $\tilde \gamma = 2 \sigma_w \sqrt{2 \log(N)}$ for such algorithm. As expected,  known radar delays result in visibly better SER, which is the reward for the additional overhead due to the transmission of pilot signals and for the increased overall complexity due to system co-ordination.

Notice that, as the SNR is 8dB, the performance gain achieved by both CS algorithms is limited: this is obviously due to the fact that, under fading channel, the performance of even an interference-free OFDM/QPSK would be poor, and the coupling between data demodulation and interference removal explains the poor performance at low SIR. For example, the considered Ricean channel with Rice factor 9 is approximately equivalent to a Nakagami-m fading with parameter $m \simeq 5.3$ \cite[p.79]{goldsmith}, whereby the error probability for an isolated OFDM/QPSK at $\text{SNR}=10$dB would be slightly larger than $10^{-2}$, a value which is approximately restored as the SIR becomes increasingly large.

As the SNR is larger than 14dB, the CS-L1 provides significant improvements over Iteration 0, but is significantly outperformed by the CS-AN especially when SIR is -5dB. This is due to the fact that, in the low SIR region, basis mismatch prevents correcting the demodulation errors in the first iteration $\bm{\hat b}^{(0)}$ through a CS-L1 algorithm, while the CS-AN algorithm, much more accurate in detecting and ranging the interference sources, allows a much more effective error correction. Needless to say, both algorithms restore the original OFDM/QPSK performance for increasingly large SIR in a much faster way than Iteration 0: it might thus be inferred that, even at SIR as low as -5dB, the SER achieved through the proposed CS-AN algorithm - in the order of  $10^{-2}$ - would in principle allow communication to be sustained once forward error correction (FEC) decoding \cite{ji2005rate} is undertaken, while no communication could take place if either of the other two algorithms were adopted.

\begin{figure*}
	\centering
	
	\subfloat[][]{\includegraphics[width=3in]{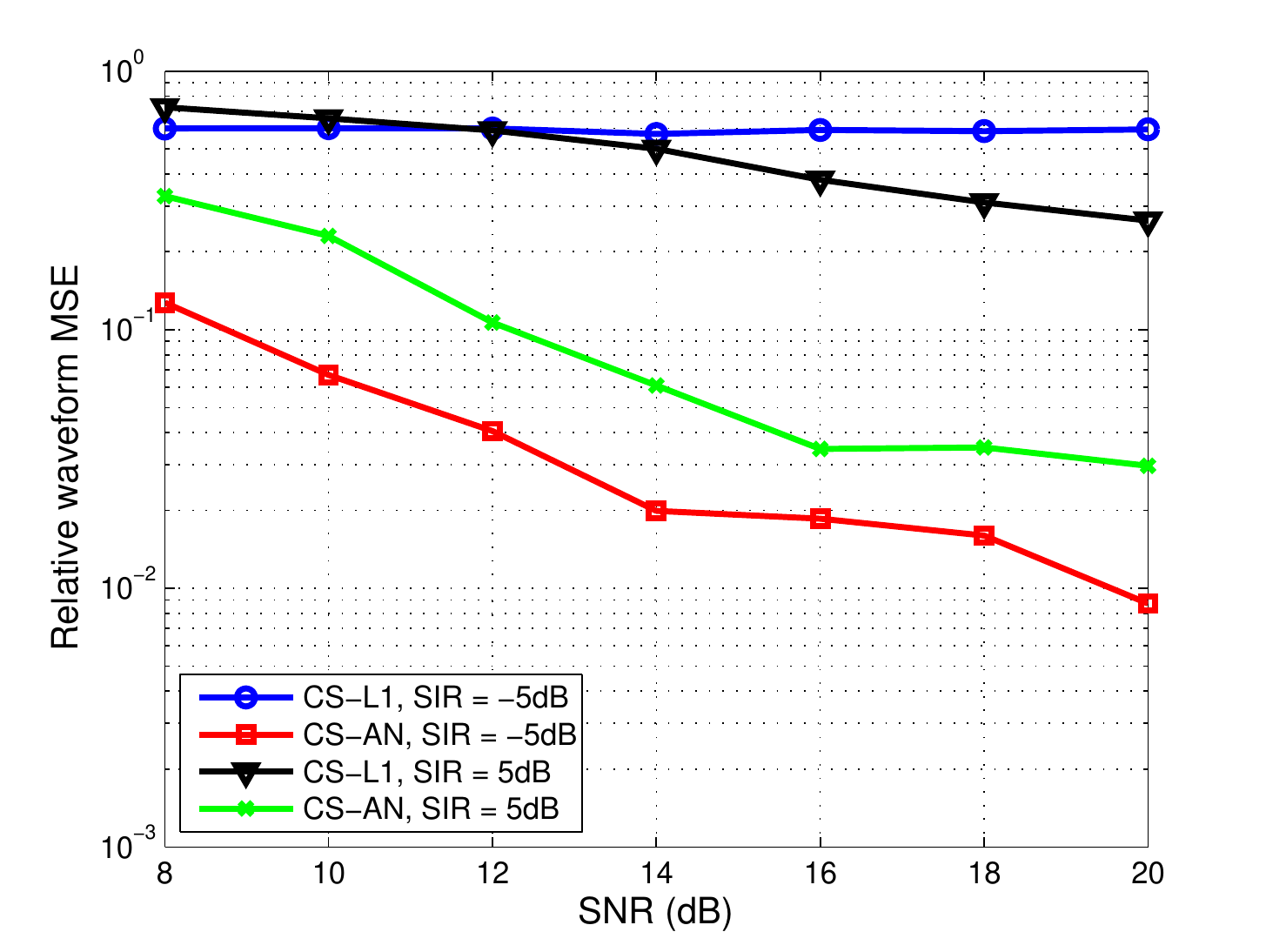}}
	\subfloat[][]{\includegraphics[width=3in]{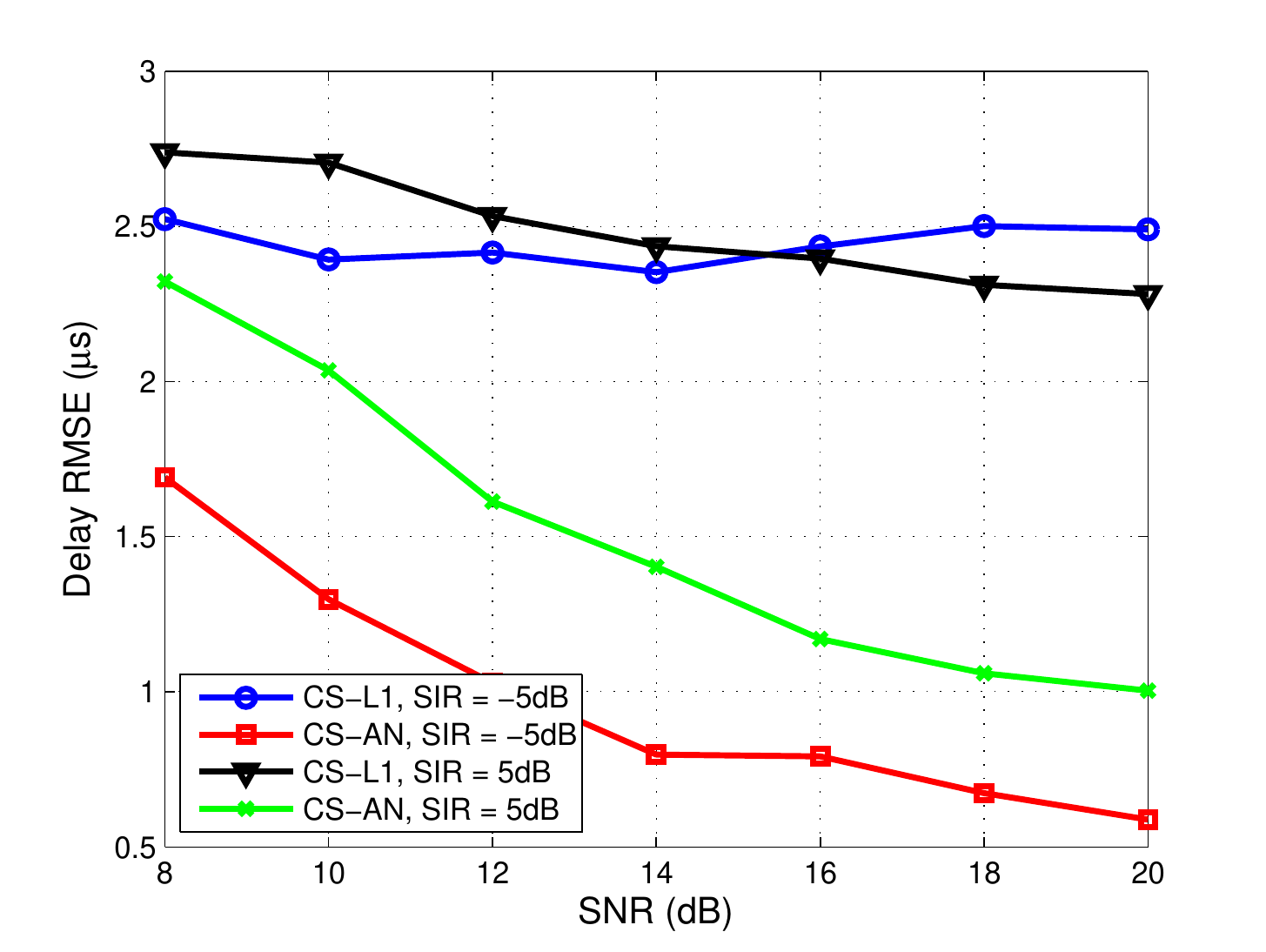}}
	
	\caption{Plots of (a) relative waveform MSE and (b) delay RMSE for different SNRs. }
	\label{fig:Accuracy}
\end{figure*}

We then evaluate the relative waveform MSE and delay RMSE of the proposed CS-L1 and CS-AN algorithms. Note that the received signal contains both communication and radar signal, and the estimation accuracy depends on both the power of the radar signals to be estimated and the performance of the demodulator: as a consequence, the estimation accuracy may be not necessarily increasing with the SIR, especially for ``intermediate" SIR values where radar signals are not strong enough to prevail on the communication signal, but still produce significant demodulation error. In Fig. \ref{fig:Accuracy}, the relative waveform MSE and delay RMSE is represented as a function of the SNR. As expected, the CS-AN algorithm provides much better accuracy than the CS-L1 algorithm. Notice also the apparently contradictory effects of the SNR. Indeed, large SNR's guarantee good demodulation performance, with a beneficial effect on the interference estimation due to the coupling. Under this point of view, the advantage of the CS-AN over the CS-L1 is visible, and the former seems to take much greater advantage of the more reliable demodulation process granted by larger values of the SNR. % On the other hand, the results also demonstrate that, as the SNR is low, the SIR has paradoxically a detrimental effect on the estimation accuracy: this is because, for large SIR and low SNR, the demodulation process is completely unreliable, which explains the trends of the curves in Fig. \ref{fig:Accuracy}.

\begin{figure*}
	\centering
	
	\subfloat[][]{\includegraphics[width=3in]{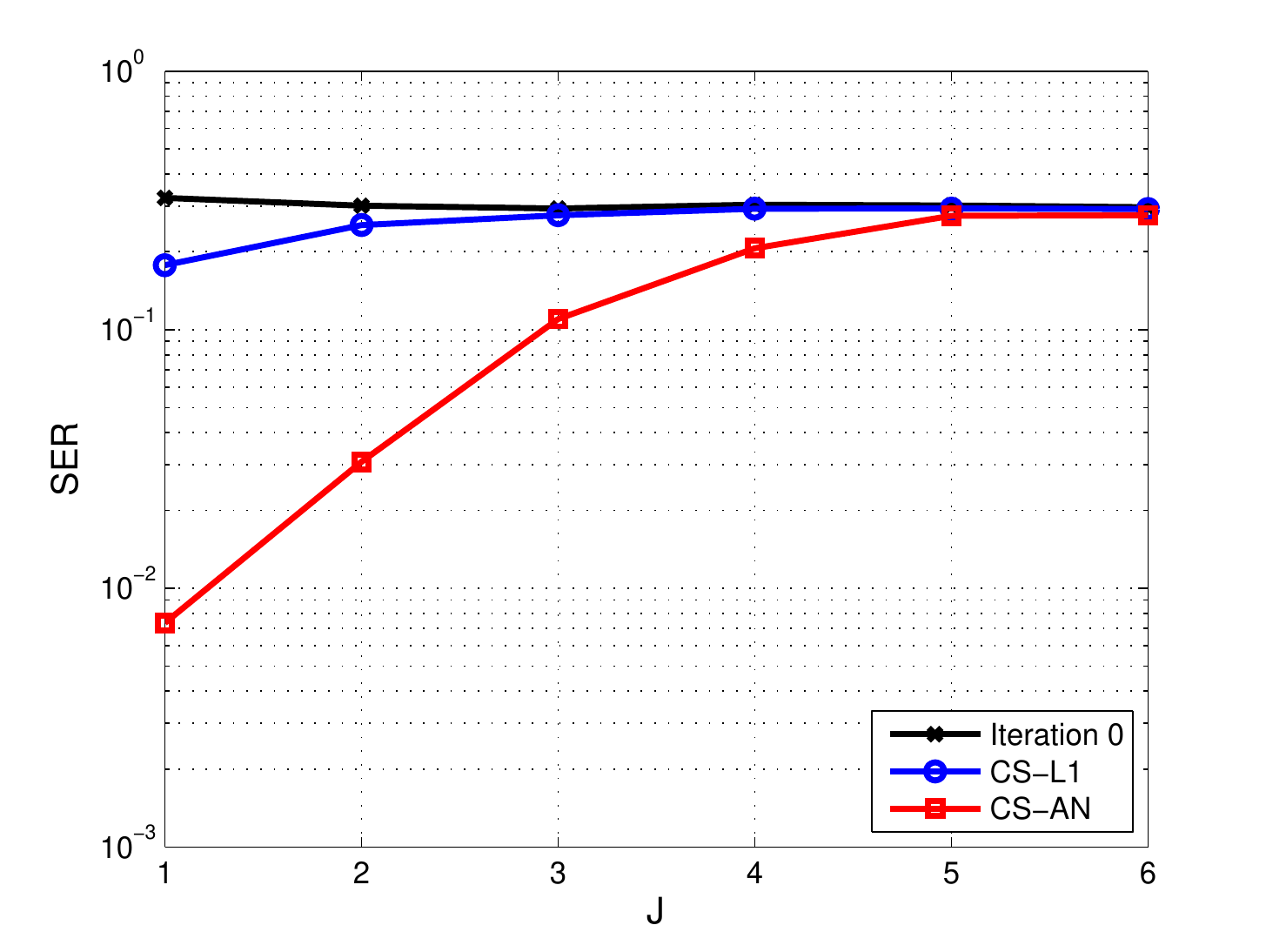}}
	\subfloat[][]{\includegraphics[width=3in]{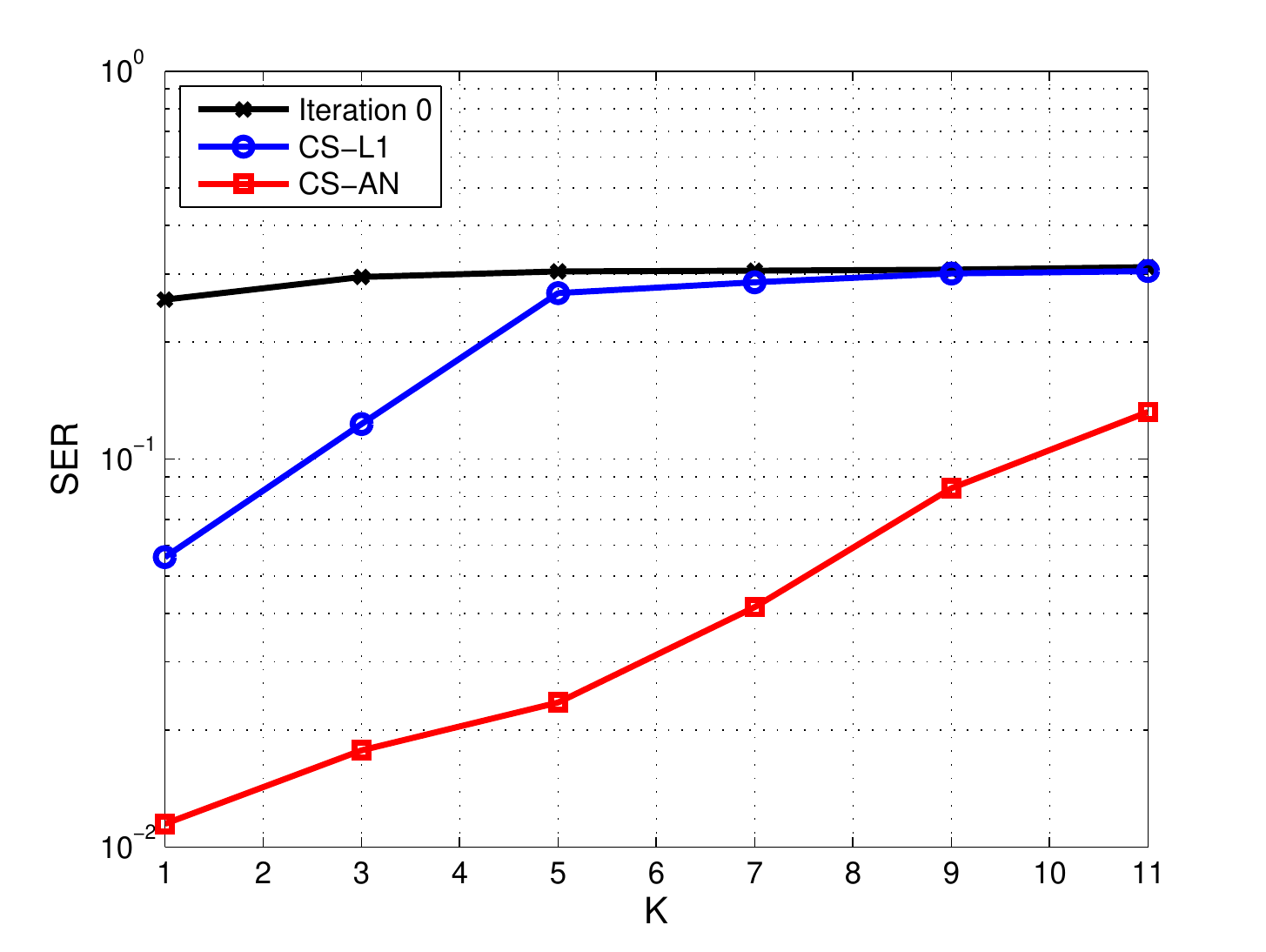}}
	
	\caption{Plots of SER against (a) $J$, and (b) $K$.}
	\label{fig:varing}
\end{figure*}

\begin{figure}
	\centering
	
	\subfloat{\includegraphics[width=3in]{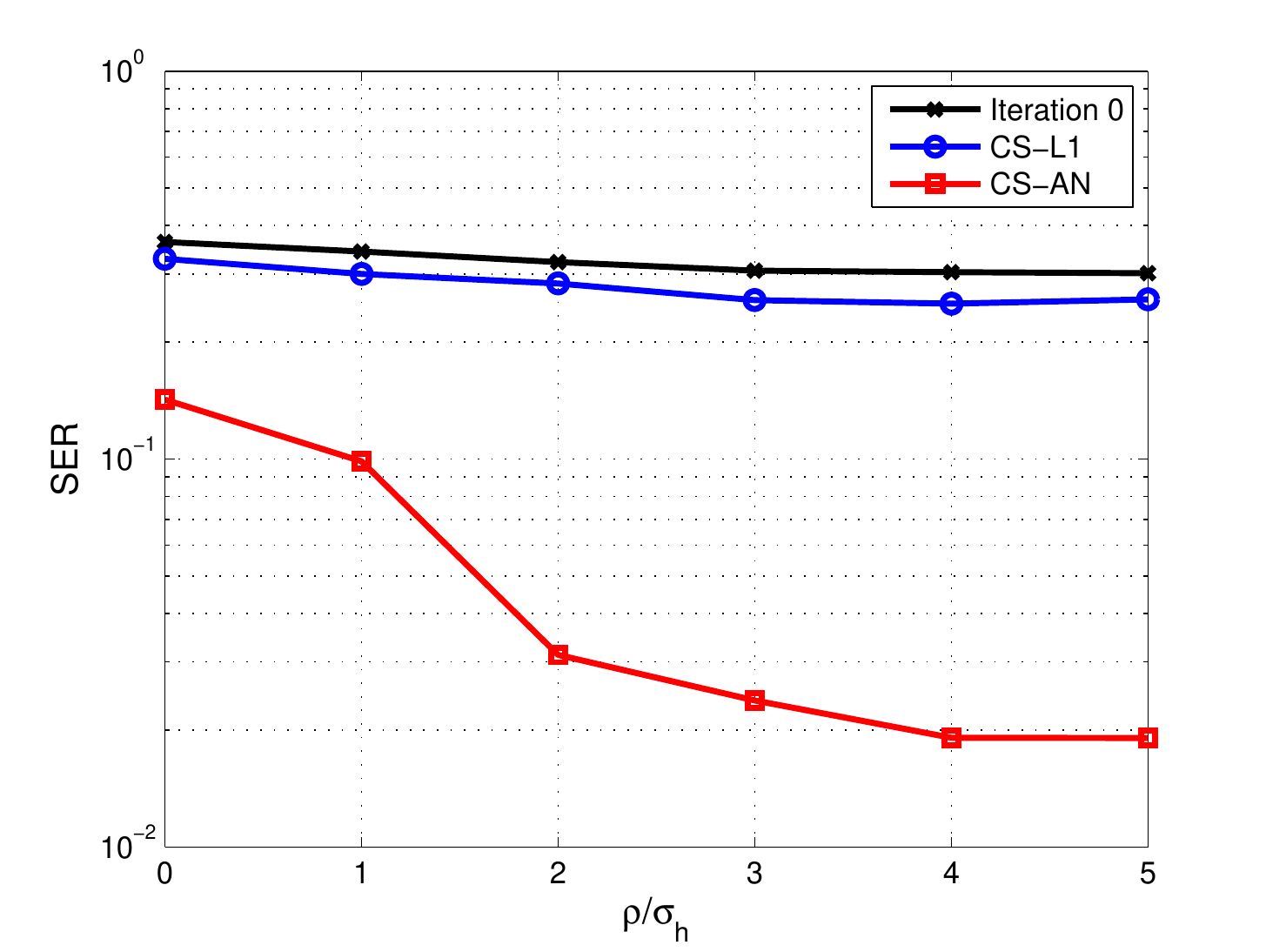}}
	
	\caption{Plots of SER against $\rho/\sigma_h$.}
	\label{fig:varingrho}
\end{figure}

The effects of $J$ and $K$ are elicited in Fig. \ref{fig:varing}. The simulations are run with an SNR of 18dB and an SIR of 0dB. In Fig. \ref{fig:varing}a, we set $\rho/\sigma_h = 3$ and plot the SER against the number of active radars: As $J$ increases, the sparsity of the problem is reduced, and the sources of interference - with the respective unknown parameters to be estimated - increase, which obviously results in a visible performance degradation for both algorithms, up to a point that not even an CS-AN algorithm is effective if $J \geq 5$. In Fig. \ref{fig:varing}b, the number of active radars is set as $J=2$ and we examine the SER behavior for varying dictionary size $K$. Although a performance degradation is evident, the robustness of the CS-AN algorithm to $K$ is a definitely appealing advantage over the other two algorithms. Finally, we investigate the effect of the channel model on the performance in Fig. \ref{fig:varingrho}, assuming $J=2$, SIR$=0$dB and SNR$=18$dB. On the horizontal axis we represent the Ricean factor, whereby decreasing values of $\rho/\sigma_h$ represent weaker and weaker direct paths, up to the case $\rho=0$ where no direct path is present. Notice that only the CS-AN algorithm guarantees, in such a severe scenario, an acceptable performance and is able to restore an error probability in the order of $10^{-2}$ as $\rho/\sigma_h>2$.

\section{Conclusions}

In this paper, we have proposed two algorithms for joint waveform estimation and demodulation in the overlaid communication and radar systems. One of them is based on the on-grid compressed sensing (CS) technique and uses $\ell_1$-norm to exploit the sparsity of the radar signal components and the sparsity of the demodulation error. The other one is a CS-based algorithm using both the atomic norm and the $\ell_1$-norm to exploit the sparsity of the radar signal components and the sparsity of the demodulation error, respectively. We have derived a fast algorithm to compute the solution to the formulated CS-AN problem. Simulation results show that the proposed algorithms provide better SER compared to the original demodulation. Future work will include extending the algorithm to the fast time-varying environment in which the mismatch in communication channel estimation and radar Doppler are considered.

\appendix

\subsection{Accuracy Analysis of \eqref{eq:barr2}}

Some approximation has been applied to \eqref{eq:barr0} before we get \eqref{eq:barr2}. In this section, the accuracy of our approximation is analyzed. As anticipated \eqref{eq:integ} holds rigorously if $0 \leq \beta \leq 1/N$ and approximately for small to moderate excess bandwidths \cite{glover2010digital}. With small $\beta$, the approximation error comes from \eqref{eq:barr} when we extend the integration from a finite range to infinite, which causes descrepency between $\bar r(k)$ and the one calculated by \eqref{eq:barr2}. For clarity, we denote $\tilde r(k)$ as the approximated $\bar r(k)$ calculated by \eqref{eq:barr2}. Let $S_{\xi} (f) = {\cal F}(R_{\xi}(t))$, then $R_{\xi}(t) = {\cal F}^{-1}(S_{\xi} (f))$, where ${\cal F}(\cdot)$ and ${\cal F}^{-1}(\cdot)$ denote the Fourier transform and inverse Fourier transform, respectively. We have
\begin{eqnarray}
\label{eq:Rxi}
R_{\xi}(t) \sim \frac{1}{t^{m+1}}, t\to +\infty,
\end{eqnarray}
where $m$ is the order of the first discontinuous derivative of $S_{\xi}(f)$. In our paper, $\xi (\cdot)$ is a SRRC with small roll-off factor $\beta$. So $S_{\xi}(f)$ is not only continuous - as all correlations - but also differentiable - whereby $m>1$ in \eqref{eq:Rxi}. In fact in our setup the correlation goes to zero as $t^{-3}$. Suppose for $t \geq \bar T$ we have ${R_\xi }(t) \leq \frac{1}{\bar T^{m_0+1}}$ with $\bar T = T' - \tau_{\min}$ and $m_0>1$. Then the approximation error can be bounded as \eqref{eq:bound}
\begin{figure*}
\begin{eqnarray}
\label{eq:bound}
\left| {\bar r(k) - \tilde r(k)} \right| &\le& \left| {\sum\limits_{n = 0}^{N-1} {x(n)\left( \begin{array}{l}
		\int_{{\tau _{\min }} - n T - T'}^{{\tau _{\max }} + (N - n -1)T + T'} {{R_\xi }(t){e^{ - \frac{{i2\pi k(t + n T)}}{{NT}}}}dt} \\
		- \int_{ - \infty }^\infty  {{R_\xi }(t){e^{ - \frac{{i2\pi k(t + n T)}}{{NT}}}}dt}
		\end{array} \right)} } \right| \nonumber \\
&& + \left| {\sum\limits_{j = 1}^J {\sum\limits_{n = 0}^{N-1} {{c_j}{g_j}(n)\left( \begin{array}{l}
			\int_{{\tau _{\min }} - n T - T' - {\tau _j}}^{{\tau _{\max }} + (N - n - 1)T + T' - {\tau _j}} {{R_\xi }(t){e^{ - \frac{{i2\pi k(t + n T + {\tau _j})}}{{NT}}}}dt} \\
			- \int_{ - \infty }^\infty  {{R_\xi }(t){e^{ - \frac{{i2\pi k(t + n T + {\tau _j})}}{{NT}}}}dt}
			\end{array} \right)} } } \right| \nonumber \\
&\le& \sum\limits_{n = 0}^{N-1} {\left| {2x(n)} \right|\int_{\bar T}^\infty  {\left| {{R_\xi }(t){e^{ - \frac{{i2\pi k(t + n T)}}{{NT}}}}} \right|dt} } \nonumber \\
&&+ \sum\limits_{j = 1}^J {\sum\limits_{n = 0}^{N-1} {\left| {2{c_j}{g_j}(n)} \right|\int_{{{\bar T}}}^\infty  {\left| {{R_\xi }(t){e^{ - \frac{{i2\pi k(t + n T + {\tau _j})}}{{NT}}}}} \right|dt} } } \nonumber \\
&\le& \sum\limits_{n = 0}^{N-1} {\left| {2x(n)} \right|\int_{\bar T}^\infty  {{t^{ - {m_0} - 1}}dt} }  + \sum\limits_{j = 1}^J {\sum\limits_{n = 0}^{N-1} {\left| {2{c_j}{g_j}(n)} \right|\int_{{{\bar T}}}^\infty  {{t^{ - {m_0} - 1}}dt} } } \nonumber \\
&\le& \sum\limits_{n = 0}^{N-1} {\frac{{2\left| {x(n)} \right|}}{{{m_0}{{\bar T}^{{m_0}}}}}}  + \sum\limits_{j = 1}^J {\sum\limits_{n = 0}^{N-1} {\frac{{2\left| {{c_j}{g_j}(n)} \right|}}{{{m_0}{{\bar T}^{{m_0}}}}}} } ,
\end{eqnarray}
\end{figure*}
where the second inequality is the result of relation
\[\bar T \leq \min_{1 \leq j\leq J, 0 \leq n \leq N-1} \left\{ \begin{array}{l}
{\tau _{\max }} - {\tau _j} + (N - n - 1)T + T',\\
T' + nT - {\tau _{\min }}
\end{array} \right\}.\]
Obviously, as $T'$ is very large, $\bar T$ becomes large so the approximation error is very small.
\color{black}

\subsection{Proof of Lemma 2}

Suppose the solution to \eqref{eq:atomicnorm} is $\bm{\hat X} = \sum_{j=1}^{\hat J} \hat c_j \bm{\hat h}_j \bm a(\hat \tau_j')^H$, the following lemma states the condition of the unique atomic decomposition:
\begin{lemma}\cite{yang2014exact}
	$\bm{\hat X}$ is the unique atomic decomposition satisfying that $\| \bm{\hat X} \|_{\cal A} = \sum_{j=1}^{\hat J} |\hat c_j|$ if $N \geq 257$ and $\Delta_{\hat \tau'} \geq \frac{1}{(N-1)/4}$.
\end{lemma}

The lemma above gives the value of $\| \bm{\hat X} \|_{\cal A}$. Since \eqref{eq:atomicnorm} and \eqref{eq:P1} are equivalent, we obtain
\begin{eqnarray}
\label{eq:atomic2}
\| \bm{\hat X} \|_{\cal A} = \mathop {\inf}\limits_{\bm{\hat U}, \bm{\hat T}} \left\{ \begin{array}{l}
\frac{1}{2N}{\rm{Tr}}(\bm{\hat U}) + \frac{1}{2} {\rm Tr}(\bm{\hat T}),\\
{\rm s.t.,} {\cal P}_{\rm Toep}(\bm{\hat U}) = \bm{\hat U}, \bm{\hat Z} \succeq 0
\end{array} \right\},
\end{eqnarray}
where the relation of $\bm{\hat Z}$, $\bm{\hat U}$ and $\bm{\hat T}$ are given in \eqref{eq:Z}. Hence, $\bm{\hat Z}$ is the solution to \eqref{eq:P1} once the equality holds. Then we need to prove that there exist Toeplitz matrix $\bm{\hat U}$ and matrix $\bm{\hat T}$ such that $\frac{1}{2N} {\text{Tr}}\left( \bm{\hat U} \right) + \frac{\text{Tr}(\bm{\hat T})}{2} = \|\bm{\hat X}\|_{\cal A}$ with $\text{rank}(\bm{\hat Z}) = \hat J$ and ${\bm{\hat Z}} \succeq 0$. Let $\bm{\hat u} = \sum_{j=1}^{\hat J} \hat c_j \bm a(\hat \tau_j')^H$. Following the Caratheodory-Toeplitz lemma \cite{caratheodory1911variabilitatsbereich,tang2013compressed}, we have
\begin{eqnarray}
\bm{\hat U} = \text{Toep}(\bm{\hat u}) = \sum_{j=1}^{\hat J} |\hat c_j| \bm a(\hat \tau_j') \bm a(\hat \tau_j')^H.
\end{eqnarray}
In such case, the matrix
\begin{eqnarray}
\bm{\hat Z} = \sum_{j=1}^{\hat J} |\hat c_j| \left[ {\begin{array}{*{20}{c}}
	e^{i \hat \varphi_j} \bm a(\hat \tau_j') \\
	\bm{\hat h}_j
	\end{array}} \right]
\left[ {\begin{array}{*{20}{c}}
	e^{- i \hat \varphi_j} \bm a(\hat \tau_j')^H & \bm{\hat h}_j^H
	\end{array}} \right] \succeq 0
\end{eqnarray}
is rank-$\hat J$, where $\hat \varphi_j$ is the phase of $\hat c_j$. Note that $\frac{1}{2N} {\text{Tr}}\left( \bm{\hat U} \right) + \frac{\text{Tr}(\bm{\hat T})}{2} = \sum_{j=1}^{\hat J} |\hat c_j| = \|\bm{\hat X}\|_{\cal A}$, $\bm{\hat Z}$ is the solution to \eqref{eq:P1}, which accomplishes the proof.

\subsection{The calculation of $\nabla_{\bm v} \tilde \zeta( \bm V \bm V^H, \bm v)$ and $\nabla_{\bm V} \tilde \zeta( \bm V \bm V^H, \bm v)$}

The gradient $\nabla_{\bm v} \tilde \zeta( \bm V \bm V^H, \bm v)$ can be directly calculated as
\begin{eqnarray}
&&\nabla_{\bm v} \tilde \zeta( \bm V \bm V^H, \bm v) = \gamma \nabla_{\bm v} \psi_\mu (\bm{v}) \nonumber \\
&& \sum_{k= -\tilde N_1}^{\tilde N_2}  {\left( \left\langle \bm f_k , \bm H \bm A \bm v \right\rangle + \left\langle \bm X , \bm{\bar d}_k \bm e_k^H \right\rangle -  z(k) \right) \bm A^H \bm H^H \bm f_k}, \nonumber \\
\end{eqnarray}
where the $m$-th element of $\nabla_{\bm v} \psi_\mu (\bm{v}) \in \mathbb{C}^{M \times 1}$ is
\begin{eqnarray}
\nabla_{v_m} \psi_\mu (\bm{v}) = \frac{\sinh(|v_m|/\mu)}{\cosh(|v_m|/\mu)} \frac{v_m}{|v_m|}.
\end{eqnarray}

Then we derive the gradient with respect to $\bm V$. Following the chain rule, we have
\begin{eqnarray}
\nabla_{\bm V} \tilde \zeta( \bm V \bm V^H, \bm v) = \left[ \nabla_{\bm Z} \tilde \zeta( \bm Z, \bm v)|_{\bm Z = \bm V \bm V^H} \right] \bm V,
\end{eqnarray}
and \eqref{eq:dzeta}, so the problem becomes calculating the gradient $\nabla_{\bm Z} \tilde \zeta( \bm Z, \bm v)$. For the convenience of our calculation, $\tilde \zeta( \bm Z, \bm v)$ is rewritten as
\begin{figure*}
\begin{eqnarray}
\label{eq:dzeta}
\tilde \zeta( \bm Z, \bm v) &=& \frac{\lambda \text{Tr}(\bm U)}{2N} + \frac{\lambda \text{Tr}(\bm T)}{2} + \sum_{k=-\tilde N_1}^{\tilde N_2} \frac{1}{2} {\left| z(k) - \left\langle \bm f_k , \bm H \bm A \bm v \right\rangle - \left\langle \bm X , \bm{\bar d}_k \bm e_k^H \right\rangle \right|^{2}} + \gamma \psi_\mu (\bm v) \nonumber \\
&& + \frac{\rho}{2}\left[ {\sum\limits_{k = 1-N}^{N - 1} {\left( {\underbrace {\bm m_k^H{\bm m_k} - \frac{1}{{N - |k|}}\bm m_k^H{\bm{I}_{N-|k|}}{{\bm{I}_{N-|k|}^H}}\bm m_k}_{\phi_k (\bm m_k)}} \right)} } \right],
\end{eqnarray}
\end{figure*}
where $\bm U$, $\bm X$, $\bm T$ are submatrices of $\bm Z$ whose structure is given in \eqref{eq:Z}, $\bm m_k \in \mathbb{C}^{(N-|k|) \times 1}$ is the $k$-th subdiagnal elements of $\bm{U}$, $\bm I_{k} = [1,1,...,1]^T$ is an $k$ dimensional all one vector. After manipulation, $\tilde \zeta( \bm Z, \bm v)$ can be rewritten in a quadratic form:
\begin{eqnarray}
\tilde \zeta( \bm Z, \bm v) = \langle \bm Z , {\cal Q}(\bm Z) \rangle /2 + \langle \bm C, \bm Z \rangle + \bar \zeta(\bm v),
\end{eqnarray}
where $\bar \zeta(\bm v)$ is a function that depends on $\bm v$; $\bm C$ and ${\cal Q}(\bm Z)$ can be respectively computed by
\begin{eqnarray}
\label{eq:C}
\bm C &=& \frac{1}{2}\left[ {\begin{array}{*{20}{c}}
	{\frac{\lambda}{N}\bm{E}}_{N}&{ - \bm{\bar Y}^H}\\
	{- {\bm{\bar Y}}}& \lambda\bm{E}_{K}
	\end{array}} \right], 
\end{eqnarray}
and \eqref{eq:Q}.
\begin{figure*}
\begin{eqnarray}
\label{eq:Q}
{\cal Q} (\bm Z) &=& \left[ {\begin{array}{*{20}{c}}
	{\bm \Xi (\bm{U})} & {\sum_{k=-\tilde N_1}^{\tilde N_2} \langle \bm e_k^H \bm{\bar d}_k , \bm X^H \rangle \bm e_k \bm{\bar d}_k^H}\\
	{\sum_{k=-\tilde N_1}^{\tilde N_2} \langle \bm{\bar d}_k \bm e_k^H, \bm X \rangle \bm{\bar d}_k \bm e_k^H}& \bm 0_{K,K}
	\end{array}} \right].
\end{eqnarray}
\end{figure*}
Here $\bm 0_{K,K} \in \mathbb{C}^{K \times K}$ is a zero matrix; $\bm E_N$ is an identity matrix with dimension $N$;
\begin{eqnarray}
\bm{\bar Y} &=& \sum_{k=-\tilde N_1}^{\tilde N_2} \bm e_k \bm{\bar d}_k^H {\left( z(k) - \left\langle \bm f_k , \bm H \bm A \bm v \right\rangle  \right)}, \\
\bm \Xi(\bm{\bar Z}) &=& \frac{\rho}{2} \sum_{k=1-N}^{N-1} \text{diag}\left(\nabla {\phi}_k (\bm m_k) ,k \right),
\end{eqnarray}
with
\begin{eqnarray}
\nabla {\phi}_k (\bm m_k) = 2\left( {\bm m_k - \frac{1}{{N - |k|}}{\bm{I}_{N-|k|}}{{\bm{I}_{N-|k|}^H}}{\bm m_k}} \right),
\end{eqnarray}
and $\text{diag}(\bm m,k)$ outputs an $N \times N$ matrix whose $k$-th sub-diagnal is the input vector $\bm m$, and the rest of the elements are zero. The derivative of $\tilde \zeta( \bm Z, \bm v)$ is
\begin{eqnarray}
\label{eq:gradZ}
\nabla_{\bm Z} \tilde \zeta( \bm Z, \bm v) = {\cal Q} (\bm Z) + \bm C.
\end{eqnarray}
Plugging \eqref{eq:gradZ} into \eqref{eq:dzeta}, $\nabla_{\bm V} \tilde \zeta( \bm V \bm V^H, \bm v)$ can be obtained.

\bibliographystyle{IEEEtran}
\bibliography{database} 

\end{document}